\documentclass[usenatbib]{mn2e}
\usepackage{natbib}  
\usepackage{epsfig}

\catcode`\@=11
\def\gta{\ifmmode{\,\mathrel{\mathpalette\@versim>\,}}
    \else{$\,\mathrel{\mathpalette\@versim>}\,$}\fi}
\def\lta{\ifmmode{\,\mathrel{\mathpalette\@versim<\,}}
    \else{$\,\mathrel{\mathpalette\@versim<}\,$}\fi}
\def\@versim#1#2{\lower 2.9truept \vbox{\baselineskip 0pt \lineskip
    0.5truept \ialign{$\m@th#1\hfil##\hfil$\crcr#2\crcr\sim\crcr}}}
\catcode`\@=12  

\def\kms{\,{\rm km}\,{\rm s}^{-1}}

\def\Gyr{\,{\rm Gyr}}
 
\def\pc{\,{\rm pc}}
\def\kpc{\,{\rm kpc}}

\def\e{{\rm e}}

\def\feh{\hbox{[Fe/H]}}
\def\cafe{[\hbox{Ca}/\hbox{Fe}]}
\def\ofe{[\hbox{O}/\hbox{Fe}]}

\def\afe{[\alpha/\hbox{Fe}]}
\def\dex{\,{\rm dex}}
\def\figref#1{Fig.~\ref{#1}}
\newcommand{\beq}{\begin{equation}}
\newcommand{\eeq}{\end{equation}}

\title[Origin and structure of the Galactic disc(s)]
{Origin and structure of the Galactic disc(s)}

\author[R. Sch\"onrich and J. Binney]{Ralph Sch\"onrich$^{1,}$$^2$\thanks{E-mail:
rasch@mpa-garching.mpg.de} and James  Binney$^3$\\
$^{1}$ Max-Planck-Institut f\"ur Astrophysik, Karl-Schwarzschild-Str. 1, D-85741 Garching, D \\
$^{2}$ Universit\"atssternwarte M\"unchen, Scheinerstr. 1, D-81679 M\"unchen, D \\
$^{3}$ Rudolf Peierls Centre for Theoretical Physics, Keble Road, Oxford OX1 3NP, UK\\}

\begin{document}

\date{Draft, February 24, 2009}

\pagerange{\pageref{firstpage}--\pageref{lastpage}} \pubyear{2009}

\maketitle

\label{firstpage}

\begin{abstract}
We examine the chemical and dynamical structure in the solar neighbourhood of
a model Galaxy that is the endpoint of a simulation of the chemical evolution
of the Milky Way in the presence of radial mixing of stars and gas. Although
the simulation's star-formation rate declines monotonically from its unique
peak and no
merger or tidal event ever takes place, the model
replicates all known properties of a thick disc, as well as matching
special features of the local stellar population such as a metal-poor extension
of the thin disc that has high rotational velocity. We divide the disc by
chemistry and relate this dissection to observationally more convenient
kinematic selection criteria. We conclude that the observed chemistry of the Galactic
disc does not provide
convincing evidence for a violent origin of the thick disc, as has been widely
claimed.

\end{abstract}


\section{Introduction}

Our Galaxy's stellar disc was first divided into two components because the
vertical density profile derived from star counts could be fitted by a
superposition of two exponentials but not by a single exponential
\citep{Gilmore83}. Further investigations revealed a thick-disc component
that was characterised by a high velocity dispersion, high $\alpha$ enrichment
and a remarkably old age.  Many authors consider the thick disc to be a relic
of a turbulent era of Galactic history in which the thick disc formed from
accreted satellites and/or a thin disc was violently heated by one or more
merger events \citep[for a discussion see e.g.][]{Reddy06}. A
violent origin  of the thick disc is strongly supported by traditional models
of chemical evolution \citep[see e.g.][]{Chiappini97}.  These models require
a period of rapid star formation early in the life of the disc,
followed by a period in which star formation effectively ceased in which the
interstellar medium could be enriched with iron by SNIa and the overall
metallicity level could be brought down by accretion of metal-poor gas.

In an earlier paper \citep[][ hereafter SB09]{SB08} we showed that when one
allows for radial mixing, which is an unavoidable consequence of spiral
structure \citep{SellwoodB,Roskar1}, a two-component disc arises naturally in
the simplest model, in which the star-formation rate (SFR) is a monotonically
declining function of time from its global maximum. In this paper we examine in greater depth the
contents of the model's solar neighbourhood, especially its
characteristic stellar populations. We identify the solar-neighbourhood
signatures of the thin and thick discs and analyse the relationship between
the chemistry and kinematics of nearby stars.  This exercise enables one to
understand better the relationship between the underlying nature of the thin
and thick discs and samples of stars that have been selected by kinematic,
chemical or spatial criteria. A better understanding of this relationship is
of considerable practical importance  because
substantial allocations of telescope time are currently committed to
spectroscopic surveys (SEGUE, RAVE, HERMES, WFMOS, Gaia) that are designed to
unravel the nature and history of the thick disc, and a clear picture will
not emerge from these surveys without a secure understanding of selection
effects.

The paper is structured as follows. In Section 2 we summarise the physics
that underlies the model and analyse its prediction for the disposition of
solar-neighbourhood stars in the $(\feh,\afe)$ plane. We identify the thin
and thick discs within this plane, and show how the kinematics and ages of
stars vary within the $(\feh,\afe)$ plane. The structures we identify seem to
have all the properties expected of the Galaxy's thin and thick discs. In
Section 3 we explore the extent to which stars can be successfully assigned
to the thin and thick discs by kinematic selection. We show that this process
cannot be clean. Finally in Section 4 we sum up and relate the
characteristics of the features we have identified to the formation history
of our model. Since the SFR in the model has been monotonically decreasing
from its global maximum,
and no merger or tidal event has ever occurred in it, we conclude that,
contrary to widespread belief, features of the Galaxy, such as the overlap in
$\feh$ of the thin and thick discs do not in fact constitute convincing
evidence for a violent origin of the thick disc.

\section{The model}\label{sec:goveqs}

\subsection{Physical inputs}

\begin{figure}
\epsfig{file=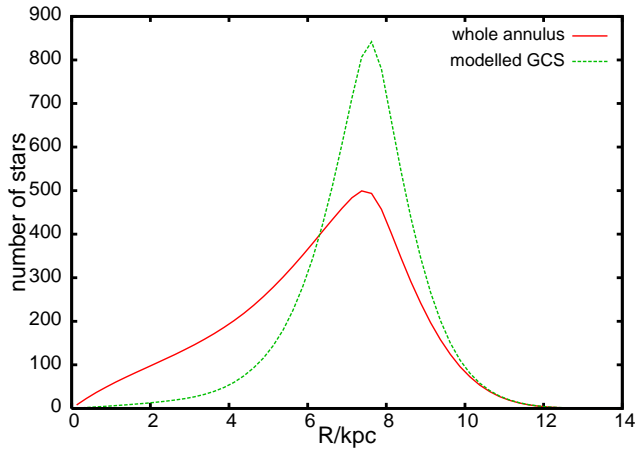,angle=-90,width=\hsize}
 \caption{The distribution of birth radii of stars in the model
   GCS sample (green dashed line) and of all stars in the solar
   cylinder (solid red line).\label{fig:origorad}}
\end{figure}

The SB09 model is the endpoint of a simulation of chemical evolution within a disc in which
the star-formation rate, which is controlled by the surface density of the
ISM as in the \cite{Kennicutt98} law, declines monotonically with time from a
unique global maximum. The gas disc always has an exponential surface density
with scale length $3.5\kpc$ so that by the Kennicutt law the young stellar disc has
an exponential surface density with scale length $2.5\kpc$. The
assumptions regarding the (universal) initial mass function, stellar
lifetimes and yields are also taken from the literature.  The only novel
features are a radial flow of gas within the disc and radial migration of
stars. The latter occurs both because over time stars change their angular
momenta (``churning'') and because they move on orbits that become
increasingly eccentric and inclined to the Galactic plane (``blurring'').
Traditional models of Galactic evolution have ignored these effects although
it has always been evident that blurring occurs. The importance of churning
was only realised when \cite{SellwoodB} found that even weak spiral structure
in N-body simulations causes stars to shift their guiding centres by a
kiloparsec and more in a single dynamical time. These motions, which arise
when a star is at the corotation resonance with a spiral arm, do not heat the
disc, so they come to light only when the angular momenta of individual stars
are followed. In the model the extent of churning is governed by a parameter
$k_{\rm ch}$ that could be determined from N-body models if we knew the past
strength of spiral structure. SB09 determined $k_{\rm ch}$ by fitting the
model to the distribution of solar-neighbourhood stars in $\feh$.  The radial
dependence of churning strength was taken to be proportional to the product
of surface density and radius $\Sigma R$, following an argument based on disc
instabilities. 

The dashed green line in \figref{fig:origorad} shows the distribution of
birth radii of stars  in the model GCS sample. Because the GCS stars lie near
the plane,  the  fraction of these stars that are  young is higher than the
fraction of young stars in the entire solar cylinder. This bias towards young stars leads
to the distribution of birth radii of GCS stars being narrower than the
distribution of birth radii for all stars in the solar cylinder, which is
show by the full red curve in \figref{fig:origorad}.
The difference between the two distributions is largest for stars born at
$\lta5\kpc$ because those inner disc stars have larger vertical velocity dispersions
and therefore larger scaleheights. 

Hot gas was assumed to be too far from the disc to take part
in churning, while the cold gas and stars were assumed to be equally involved
in this process. It is likely that these assumptions exaggerate the impact of
churning on old stellar populations, which have high velocity dispersions,
relative to its impact on young stellar populations. Since there is as yet no
basis for quantifying the impact of velocity dispersion on churning rate, the
model of Paper I avoids additional undetermined parameters by ignoring this
possibility.

\begin{figure}
  \epsfig{file=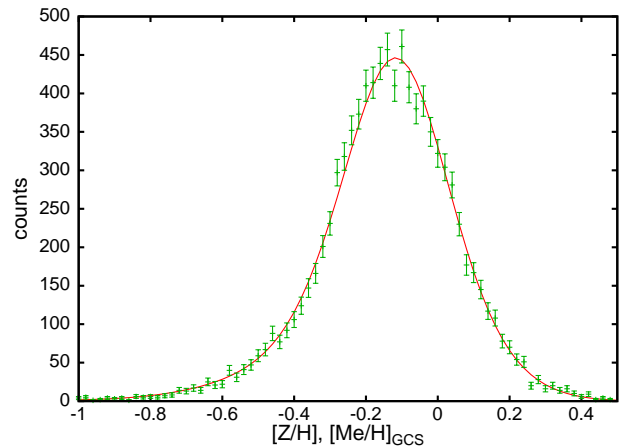,angle=-90,width=\hsize}
\caption{The metallicity distribution of solar-neighbourhood stars: data
points from Holmberg et al.\ (2007); red curve the SB09 model. For the model
total metal abundance is plotted horizontally, while for the data the plotted
quantity is the photometric metallicity indicator given in Holmberg et al.\ (2007).}\label{fig:GCS}
\end{figure}

The flow of gas within the disc enables the surface density of star formation
to be an exponential function of radius even though the rate at which
accreted gas joins the disc does not necessarily vary exponentially with
radius. The surface density of inflow of metal-poor gas and the flow of gas
within the disc are jointly controlled by two parameters, $f_{\rm A}$ and
$f_{\rm B}$, which substitute for a knowledge of the radial profile of cosmic infall.
Attempts to obtain the latter from simulations \citep[e.g.,][]{Colavitti08} have
not been successful, probably because much of the gas that joins the disc
spends a significant time after infall in the warm-hot intergalactic medium
(WHIM). Hence at the present time we must parametrise the infall in some way.
SB09 found that $f_{\rm B}$ is effectively set by the measured oxygen
gradient in the ISM, and $f_{\rm A}$ was fitted alongside $k_{\rm ch}$ to the
local metallicity distribution of stars. \figref{fig:GCS} shows the fit that
was obtained to data for $\sim10\,000$ non-binary stars in the Geneva--Copenhagen
survey \cite[][ hereafter GCS]{Nordstrom04,HolmbergNA}.

The random velocities of stars formed at a given radius are assumed to
increase with age $\tau$ as $\tau^{1/3}$ in line with the predictions of both
theory \citep{Jenkins92} and studies of the solar neighbourhood
\citep{JustJahreiss07,AumerB08}. We have modified the \cite{SB08} model very slightly by
increasing the $\langle v_R^2\rangle^{1/2}$ of a $10\Gyr$-old population of
local stars from $38\kms$ \citep[cf][]{DehnenB} to $45\kms$. This increase
brings the $\langle v_R^2\rangle^{1/2}$ for the entire GCS sample into line
with the observed value. In concordance with observations
\citep[e.g.][]{LewisFreeman89} the square of the intrinsic velocity dispersion
at a given age is assumed to scale with radius as $\e^{-R/1.5R_*}$, where
$R_*=2.5\kpc$ is the scale-length of the stellar disc, so $\langle
v_R^2\rangle^{1/2}\simeq90\kms$ at $R=R_*$. The square of the vertical
velocity dispersion component is assumed to show a slightly steeper rise
(implying approximately constant scaleheight) being proportional to $\e^{-R/R_*}$.

\begin{figure}
\epsfig{file=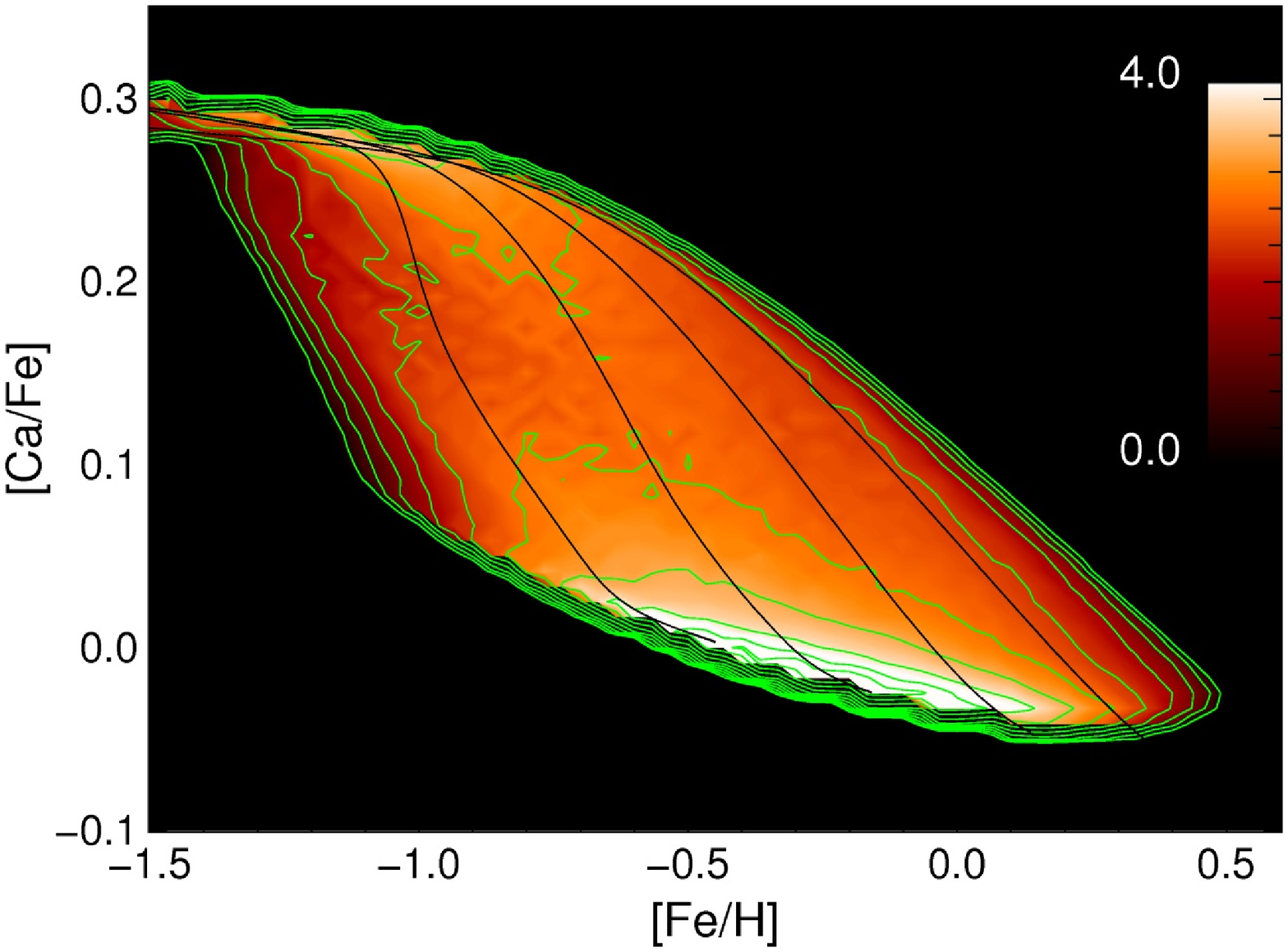,width=\hsize}
\epsfig{file=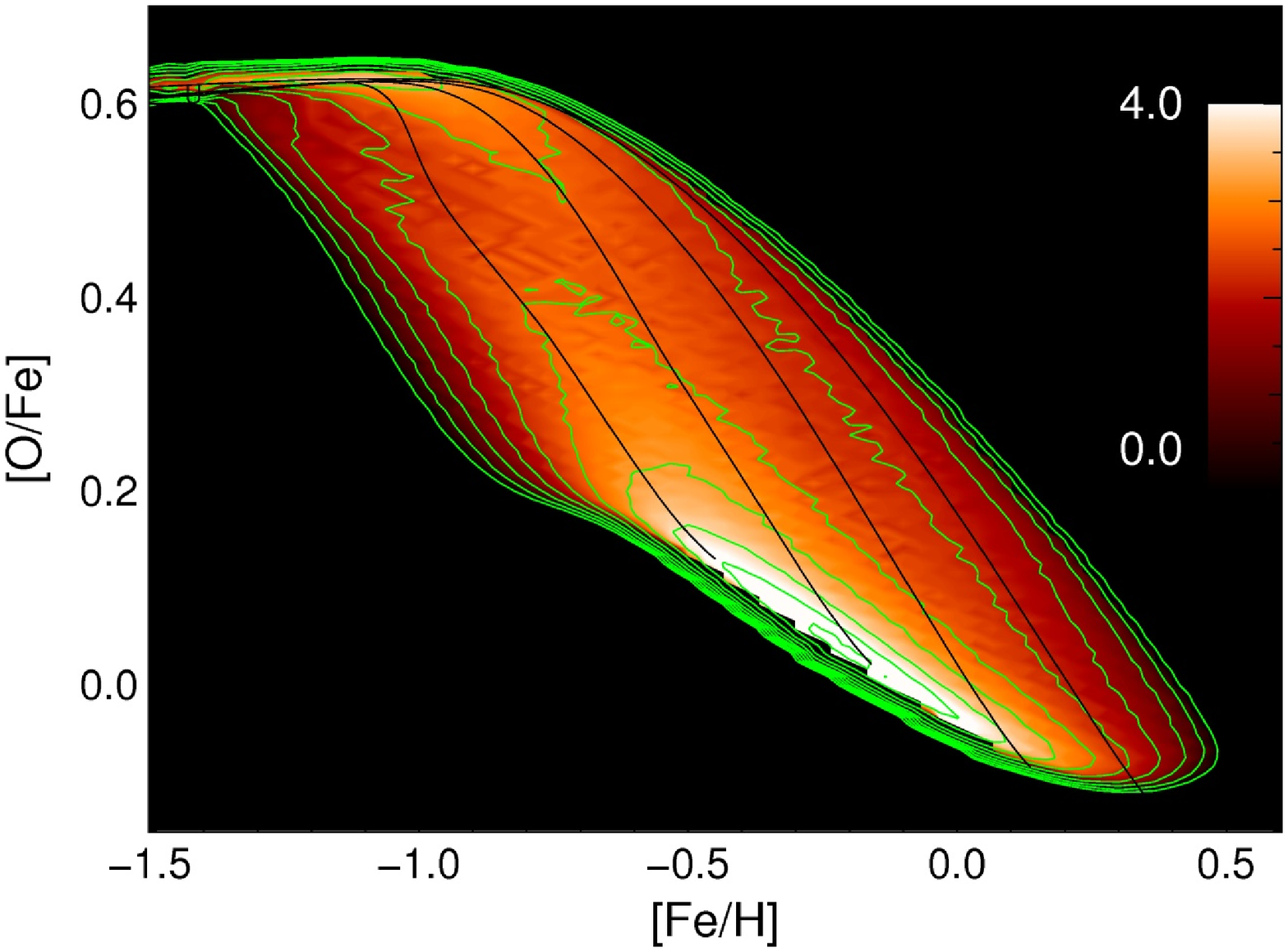,width=\hsize}
 \caption{Logarithmic stellar densities for a simulated GCS stellar sample in
the $(\cafe, \feh)$ (top) and $(\ofe, \feh)$ planes. Contours have a $0.5
\dex$ spacing. Black lines track the development of the cold ISM in annuli of
radii of (from right to left) $2.5, 5.0, 7.5$ and $10 \kpc$.\label{fig:cafe}}
\end{figure}

\subsection{Predictions of the model}

The merit of the SB09 model is that it tracks the kinematics of stars in
addition to their chemistry.  Observations always have a kinematic bias of
some kind, either because they are restricted to stars that lie near the Sun
and therefore the plane,
a region favoured by stars with small vertical velocity dispersions, or
\citep[as in][]{Juric08,Ivezic08} because they focus on faint stars that are
far from the plane, or because an explicit high-velocity criterion is applied
in order to reduce contamination of a thick-disc sample by thin-disc stars.
A model that includes both chemistry and kinematics is essential for the
interpretation of a kinematically selected study.

The SB09 model makes predictions for the global structure of the Galaxy's
stellar and gas discs, but in this paper we focus on the solar neighbourhood,
and especially the stars that happen to lie within $100\pc$ of the Sun. This
volume is of particular interest because within it G-dwarfs are
bright enough to have good Hipparcos parallaxes, measured radial velocities,
and in a few cases medium-to-high resolution spectra from which detailed
chemical abundance patterns can be extracted. The GCS provides space
velocities, surface gravities and metallicities for this volume, and detailed
abundance analyses have been carried out for much smaller subsets of stars
\citep{Fuhrmann98,Bensby03,Venn04,Bensby05,Gilli06,Reddy06}.  Because the GCS sample
is essentially magnitude limited, it is not representative of the volume
typically simulated by models of chemical evolution, namely a cylindrical
annulus around the disc.  In particular, thick-disc stars are
under-represented within the GCS relative to a cylindrical annulus. The SB09
model provides an arena in which the impact of this bias can be assessed.

\figref{fig:cafe} shows the densities of simulated GCS stars in the
$(\feh,\cafe)$ (upper) and $(\feh, \ofe)$ planes.  Trajectories of the cold ISM
at galactocentric radii of $10, 7.5, 5, 2.5 \kpc$ (from left to right) are
indicated by black lines. The ISM starts at early times with low metallicity
and high $\alpha$ enhancement at the top left of each panel. As the gas is
enriched with metals, each trajectory moves to the right.  With the onset of
SNIa, the composition of stellar yields shifts towards iron, so $\afe$
decreases and the trajectories move downwards. Eventually the ISM
approaches a steady state in which additional enrichment is balanced by the
infall of fresh metal-poor material from the IGM. Since the delay of
SNIa-enrichment is assumed to be independent of environment, the time at
which trajectories first move downward is independent of radius. By the fact
that the timescale of SNIa-enrichment does not vary with radius either, the
ISM trajectories tend to be nearly aligned. Thus the
point of turndown is at higher metallicities (further to the right) for
populations closer to the Galactic centre, where the ISM is
enriched faster by more intense star formation relative to the
present gas mass.

The colours and green contours in \figref{fig:cafe} show the density of stars
within each plane.  In each panel two ridges of high density are apparent --
one at high $\afe$, which we call the metal-poor thick disc and one at low
$\afe$, which is associated with the thin disc. Crucially, the thin-disc
ridge runs at a large angle to the black trajectories of the ISM. Thus the
thin-disc ridge in no sense traces the chemical history of the thin disc;
instead it reflects the spread in radii of birth of local stars, which gives
rise to a spread in $\feh$ by virtue of the metallicity gradient within the
ISM (which is larger in the SB09 model than in traditional models of chemical
evolution). In a similar manner the  thick-disk ridge follows the
evolution of all rings at low metallicities, but stretches significantly to
higher $\feh$ than the point at which the solar annulus leaves it.

The depression in the stellar density between the ridges in \figref{fig:cafe}
is a consequence of the rapid downward motion of all trajectories after the
onset of SNIa, and of the ISM approaching a steady state as it enters the
thin-disc ridge line; relatively few stars are formed at intermediate values
of $\afe$. Variations in the timescales of enrichment will change the depth
of the depression -- for example, a shorter timescale for the decay of SNIa
progenitors will cause trajectories to move downwards faster, leading to
fewer stars in the intermediate region. Our models use a prescription for
SNIa that is standard for models of chemical evolution, with no SNIa until
$0.15 \Gyr$ after star formation, and then an exponential decay in the rate
of SNIa with time constant $1.5 \Gyr$.  \cite{Mannucci06} suggest that
$\sim$half SNIa explode promptly (within $0.1\Gyr$ of star formation) and the
rest explode at a rate that declines exponentially with time constant
$3\Gyr$. The existence of prompt SNIa would not materially affect our work as
long as a significant fraction of SNIa are in the population with a long time
constant, since the prompt SNIa will lower the alpha-enhancement level of the
thick disc component, but not affect the evolution between the two density
ridges in the abundance plane.  \cite{Foerster06} showed that timescales are
very weakly constrained by SNIa counts due to uncertainties in the
star-formation histories. 

Since both Ca and O are $\alpha$-elements, the distributions in the upper and
lower panels of \figref{fig:cafe} are qualitatively similar, and  O will
be the only $\alpha$-element explicitly discussed below.  

\begin{figure}
\epsfig{file=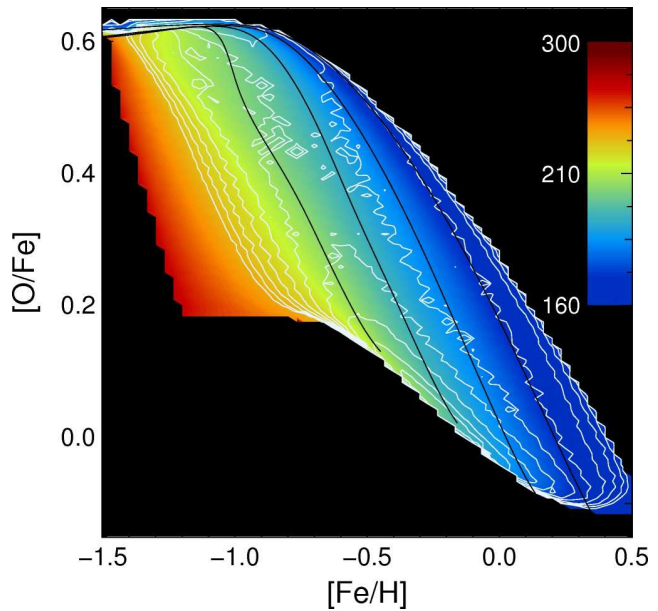,width=\hsize}
 \caption{The structure of a simulated sample of GCS stars in the $(\feh,
\ofe)$ plane. Contours spaced by $1\dex$ give the density of stars in this
plane, while colour codes the average rotational velocity of the stars found
at the point in question -- the local circular speed is assumed to be
$220\kms$. Black lines give the trajectories of the cold ISM during the model
Galaxy's evolution for galactocentric distances of (from left to right) $10,
7.5, 5, 2.5 \kpc$.\label{fig:veltra}}
\end{figure}

As \cite{Haywood08} has pointed out, the principal tracers of radial mixing
are the large dispersion in the metallicities of stars in the solar
neighbourhood and the strong increase in this dispersion with age, which is
caused by immigration of high-metallicity stars from inwards and
low-metallicity stars from outwards. In fact, as SB09 demonstrated, it is
impossible to fit the shape of the local metallicity
distribution under plausible assumptions without radial mixing. 

As well as generating a large dispersion in the metallicities of old stars,
radial migration has a big impact on the interdependence of kinematics and
chemistry. This impact is illustrated by \figref{fig:veltra}, which is
another plot of the $(\feh, \ofe)$ plane, but now with colour indicating the
mean rotation velocity of stars at each point -- blue indicates low rotation
velocities and red large ones. We see that at any given value of $\afe$,
there is a tight correlation between $\feh$ and rotation velocity in the
sense that high $\feh$ implies low rotation velocity because stars with high
$\feh$ are migrants from small radii and tend to be deficient in angular
momentum, while stars with low $\feh$ are migrants from large radii. The
black lines that show the trajectories of the ISM almost constitute contours
of constant mean rotation velocity, but there is a barely perceptible
tendency for the rotation velocity to decrease as one moves up along a black
line.

While the correlation between $\feh$ and rotation velocity seen in
\figref{fig:veltra} is qualitative consistent with stars being scattered to
more eccentric orbits while retaining their angular momenta (``blurring''),
quantitatively changes in angular momentum (``churning'') play a big role in
structuring \figref{fig:veltra}.  While churning does occasionally move the
guiding-centre radius of a star from $R_{\rm g}<R_0$ to $R_{\rm g}>R_0$ and
thus increase the mean rotation speed at large $\feh$, the dominant effect of
churning is to move guiding centres from $R_{\rm g}\ll R_0$ to $R_{\rm
g}<R_0$ such that a {\it very\/} metal-rich star is found in the solar
neighbourhood at a relatively low rotation velocity. To illustrate the impact
of churning quantitatively, if angular momentum were conserved, the
population born $5 \kpc$ from the Galactic centre would have $v_\phi\sim150
\kms$, while stars born at $10\kpc$ would have $v_\phi\sim300\kms$. In the
simulated sample, the mean speeds associated with these radii of birth
are actually $200 \kms$ and $240\kms$. 

Churning substantially increases the chemical heterogeneity of the stars that
one finds near the Sun with a given $V$ velocity: if high-$\alpha$,
high-$\feh$ stars were brought to the Sun only by blurring, then all stars
with a given $V$ and therefor angular momentum would have identical
chemistry. By changing the angular momenta of stars, churning ensures that
stars of a given chemical composition are seen near the Sun over the whole
range in $V$.

In \figref{fig:veltra} the density of stars is indicated by white contours,
which are spaced by $1 \dex$. The very top edge of the populated region is
shaded blue, indicating low rotation velocities. Thus the highest-$\alpha$
stars form a structure with a large asymmetric drift. This fact reflects the
large velocity dispersion of these stars, and the increased opportunity for
migration enjoyed by this old population. Note that in this region of the
diagram the colour rapidly changes to red as one moves downwards. Hence there
is also a population of high-$\alpha$ stars that have large rotation
velocities. These are typically slightly less old stars that formed outside
the solar radius.  The
thin-disc ridge line at low $\alpha$ enhancement ranges from the strongly
trailing at high metallicities to high rotational velocities and slight
$\alpha$ enhancement at low $\feh$, as found observationally by
\cite{Haywood08}.  

\begin{figure}
\epsfig{file=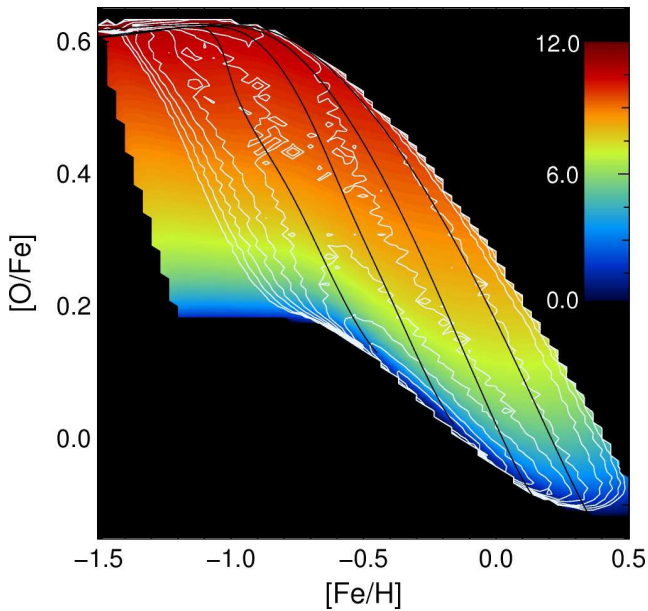,width=\hsize}
 \caption{Same as \figref{fig:veltra} with colour coding for age and $0.5
\dex$ spacing for density contours.\label{fig:velage}}
\end{figure}

In \figref{fig:velage} the contours show stellar density in increments of
$0.5\dex$ and colours encode the mean age of stars, with blue implying youth.
Naturally the oldest stars are high up, in the region of high $\afe$. Right at
the top, lines of constant age run almost horizontally. As one descends the
diagram, lines of constant age slope more steeply down to the right as a
result of the more rapid decline in $\afe$ at small radii.  The youngest
stars both from outer and from inner rings have yet to reach the solar
neighbourhood in significant numbers, so in \figref{fig:velage} several white
contours are crossed as one moves along the thin-disc ridge from the location
of the solar-radius ISM.

\begin{figure}
\epsfig{file=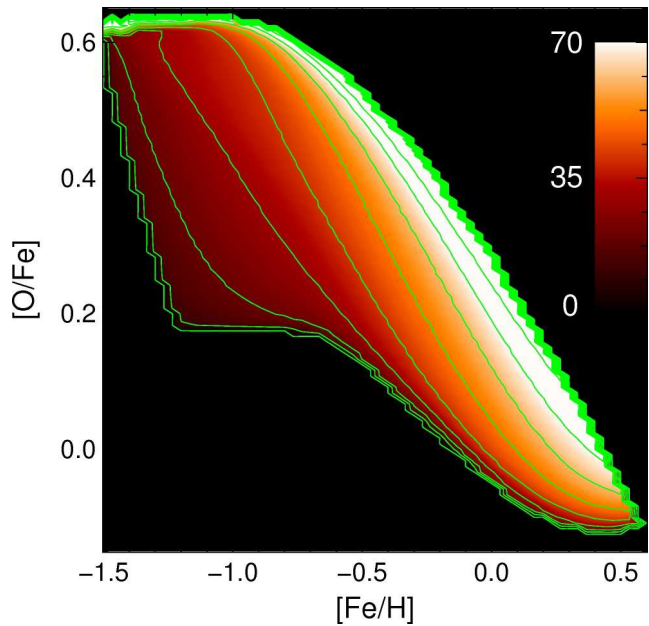,width=\hsize}
 \caption{Velocity dispersions (in $\kms$) as functions of position
in the $(\feh, \ofe)$ plane. The graph is derived for a solar-neighbourhood
sample by measuring the velocity dispersions of the populations with a
specific chemical fingerprint. Two effects act on the velocity dispersion:
The dependence on age mostly induces a top-down gradient following the
evolution lines of the ISM. In the perpendicular direction
(left-right) velocity dispersion increases with decreasing galactocentric
radius. The low dispersion of the Galactic thin disc is visible on the lower
left side, girded by a high dispersion band running from top left to bottom
right.\label{fig:veld}}
\end{figure}

In \figref{fig:veld} both colours and contours ($10\kms$ spacing) show the
velocity dispersion $\sigma_U$ of stars of a given chemical composition. The
structure of the figure is the product of two mechanisms: (i) The velocity
dispersion of stars born at any given radius scales with the third power of
age, so older stars have larger velocity dispersions than younger stars born
at the same locations.  Consequently, in the figure velocity dispersion tends
to increase from bottom to top. (ii) Velocity dispersion increases inwards,
so stars that have reached the solar neighbourhood from small radii of birth
have larger velocity dispersions than stars that have reached us from large
radii.  When this fact is combined with the fact that for any given date of
birth more metal-rich stars are born at smaller radii, a steep rise in
$\sigma_U$ from left to right arises in \figref{fig:veld}. This plot suggests
that we should include the region of high velocity dispersion along the upper
right part of the populated region in the thick disc.

\begin{figure}
\epsfig{file=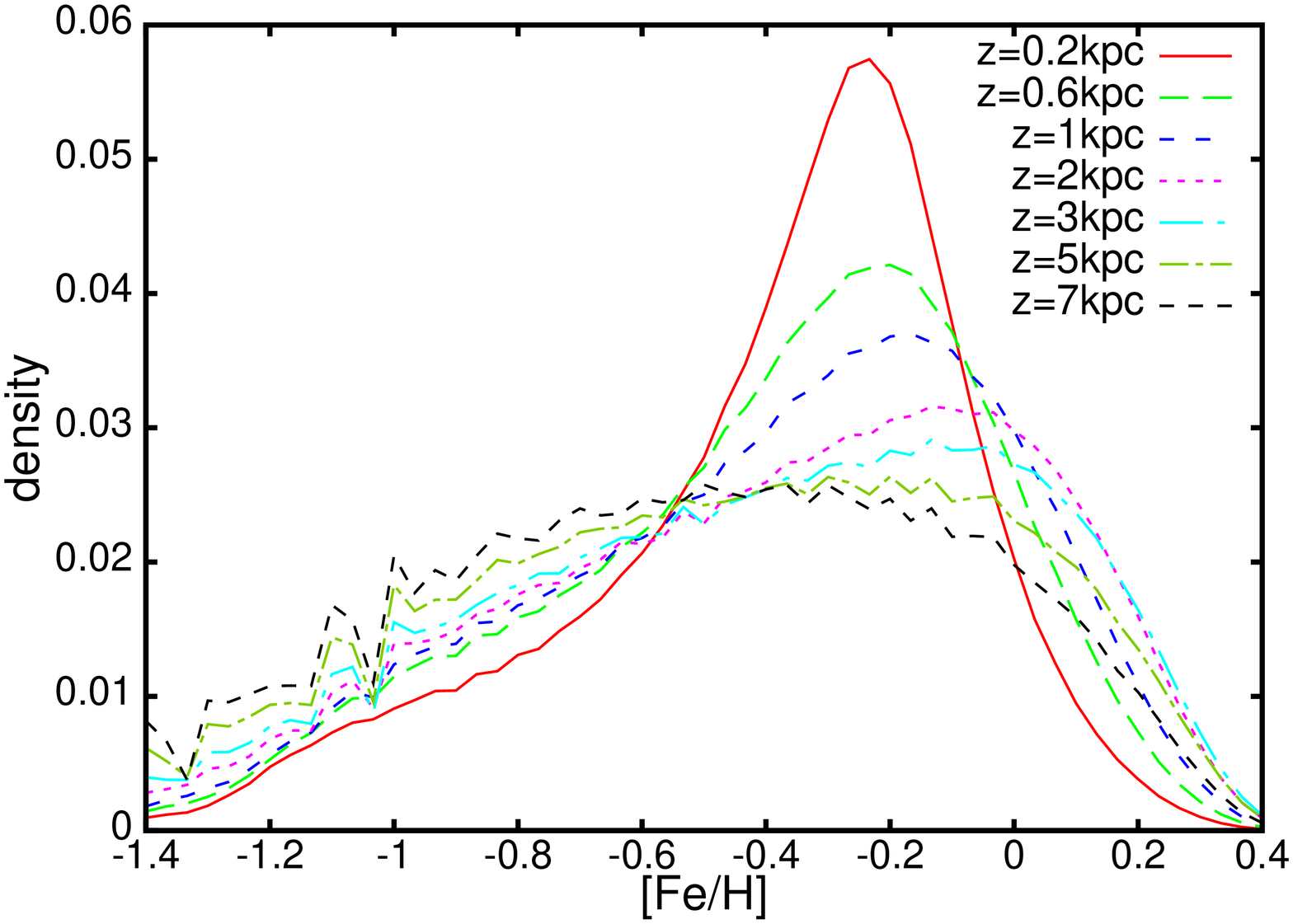,angle=-90,width=\hsize}
\epsfig{file=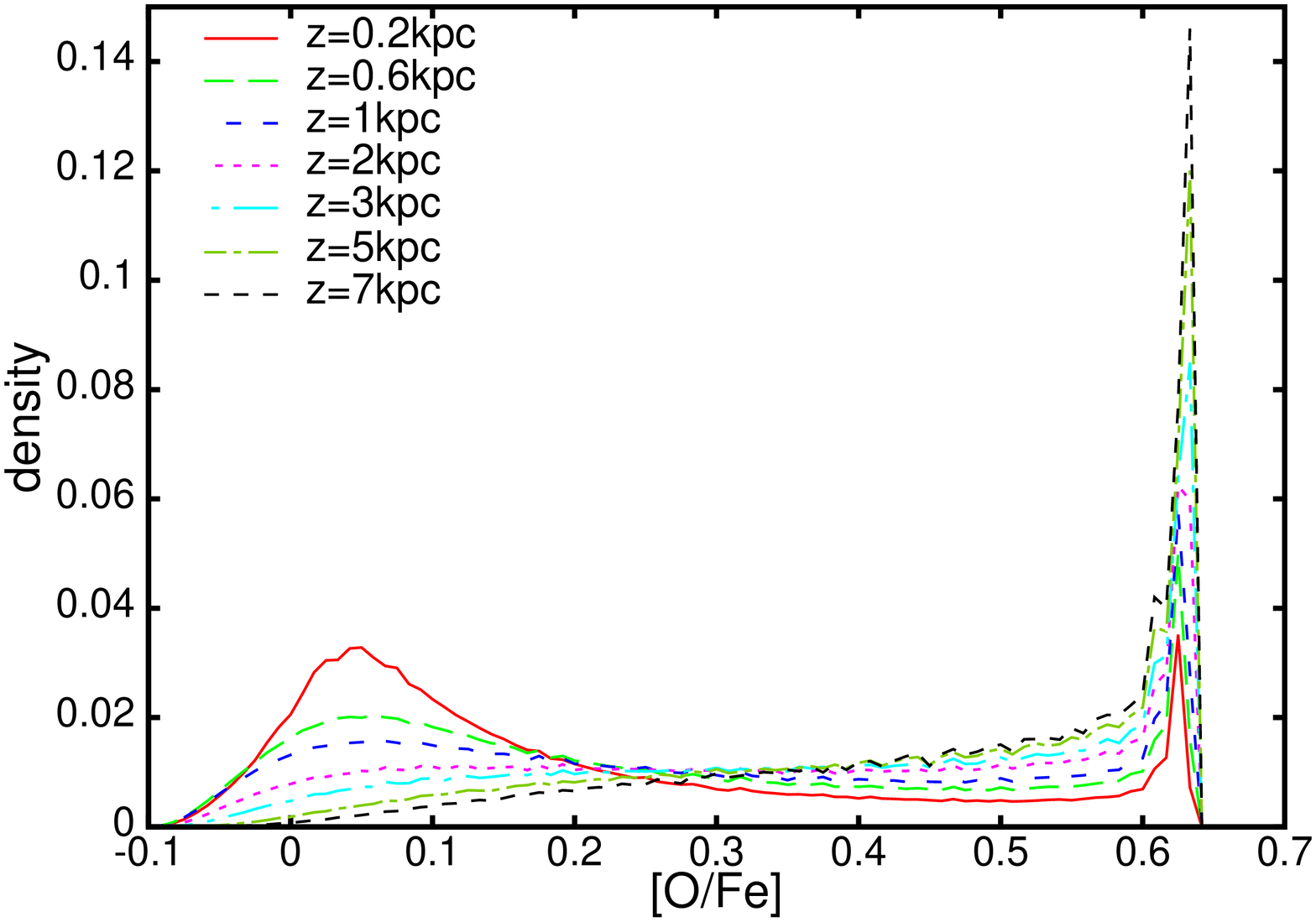,angle=-90,width=\hsize}
 \caption{The model's stellar metallicity distributions at different
   heights above the plane at $R_0$. Here we
   avoided implying a specific selection function by using the mass of
   a specific population to determine its weight in the
   distribution. The diagrams are unsmoothed and the scatter comes
   from the radial ($0.25\kpc$) and temporal ($30\,$Myr) resolution of
   the model. Upper panel: distributions of iron abundance. Lower panel:
   distributions of relative oxygen abundances. \label{fig:metheights}}
\end{figure}

Since the single populations have -- according to their ages and places of
birth -- different vertical dispersions, the older populations and those
coming from inner radii will have higher scale heights and so reduced weights
in a local sample. However, these populations dominate the composition high
above the Galactic plane. The upper panel of \figref{fig:metheights} depicts
the iron abundance distributions of the stars at different heights above the
plane. Both tails of the distribution are strengthened as one moves away from
the plane. The growth in the proportion of metal-rich stars with $|z|$ is at
first unexpected, but is a natural consequence of the higher vertical
velocity dispersion of stars in the inner disc. Notwithstanding the growth of
the metal-rich wing of the metallicity distribution, the mean metallicity
falls with increasing $|z|$ by more than $0.2 \dex$, while the dispersion
increases from below $0.3\dex$ to $0.5\dex$. We expect the model, however, to
underestimate the vertical metallicity gradient on account of our
assumption that a star inherits the velocity dispersion of the galactocentric
radius at which it was born. A better model would take account of the actual migration
paths of stars -- how long each star spent with its guiding centre at each
radius. It would predict smaller scale heights for populations of stars born
in the inner disc. Thus the model might
predict too high a fraction of high-metallicity stars to high altitudes. 

The metallicity distribution at high altitudes depends on
the weakly constrained early evolution of the disc and on details of
mixing,  so comparisons with observational data would provide valuable
constraints on these less secure aspects of the model. Unfortunately, such
comparisons are not feasible at present. In particular, we cannot compare with
the SDSS data of \cite{Ivezic08} because their metallicity determination
breaks down above $\feh \sim -0.5 \dex$. The model however
has a constant mean rotational velocity in the metallicity range
probed by the SDSS survey, in line with the data of \cite{Ivezic08}.

The lower panel of \figref{fig:metheights} shows the $\ofe$ distributions at
different heights. It reveals the bimodal structure that motivates the
division of the disc into two. The exact shape of the two peaks as well as
the number of stars in between depend on assumptions about gas enrichment and
the behaviour of SNIa, but the bimodality of the distribution is a
fundamental prediction of the model, as was shown in the appendix of SB09.
The increasing bias to high ages as $|z|$ increases is reflected in the
growing strength of the high $\ofe$ peak relative to the low $\ofe$ peak
associated with the thin disc.

\begin{figure}
\epsfig{file=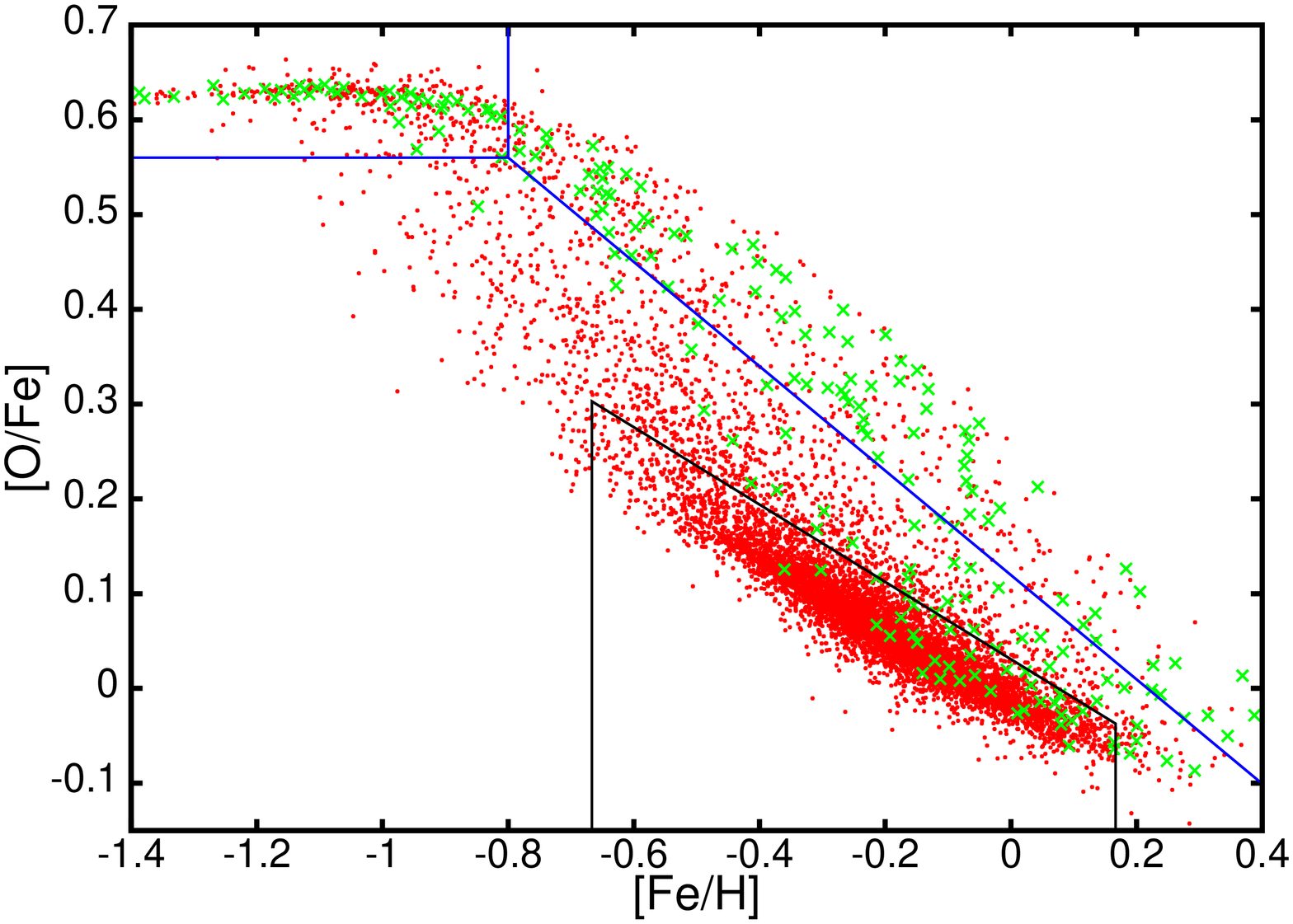,angle=-90,width=\hsize}
\epsfig{file=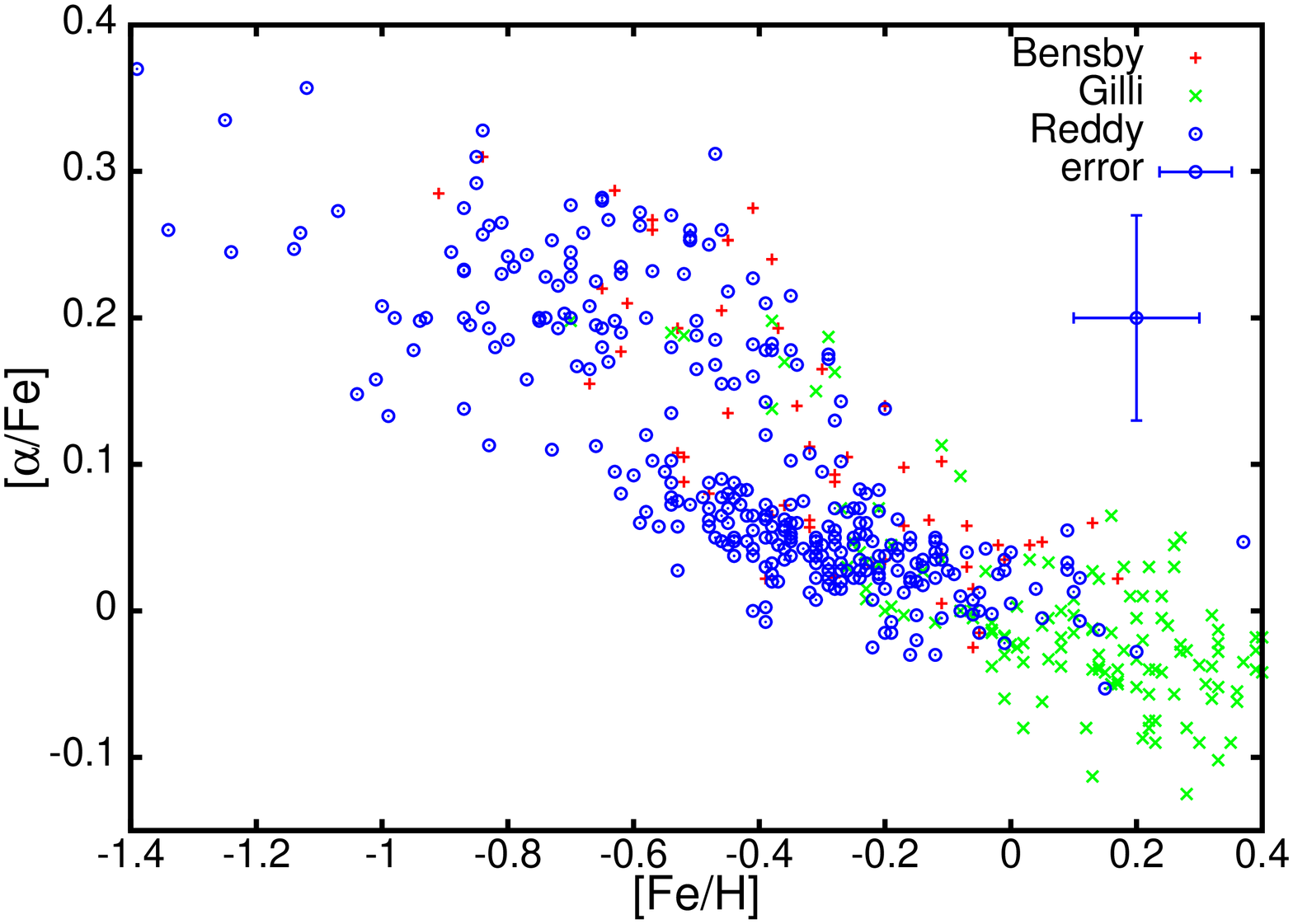,angle=-90,width=\hsize}
 \caption{Upper panel: a scatter plot for a GCS-like measurement of
solar-neighbourhood stars in the $(\feh, \ofe)$ plane. Red dots mark
positions of stars, while green crosses mark stars that are selected to the
thick disc via the kinematic selection scheme. Lines mark possible criteria
to dissect the data with a chemical classification scheme in the $(\feh,
\ofe)$ plane. Lower panel: the locations in the $(\feh,\afe)$ plane of stars
with spectroscopically determined chemical compositions from Bensby (2005),
Gilli (2006) and Reddy (2006).\label{fig:tcuts}}
\end{figure}

\subsection{The disc divided}

There are principally two strategies by which the disc has classically been
dissected: by kinematics and by chemistry. We caution that different
selection procedures do yield intrinsically different samples and that in
general these are not equivalent. We shall see that these selection
differences, which account for the spread by almost an order of magnitude in
estimates of the relative local densities of the thick and thin discs, are
readily understood in the context of our model. In each scheme criteria are
set that define both thin and thick disc components, while stars that meet neither
criterion are here assigned to an ``intermediate population''. We turn first to
chemical selection and then in the light of this assess the quality and
effects of kinematical criteria.  

The dots and crosses in the upper panel of
\figref{fig:tcuts} show a realisation of a GCS-like sample of stars in the
model. The ridge of the thin disc is evident, as is a ridge of metal-poor
thick-disc stars at $\feh\lta-0.65$ and $\ofe\sim0.6$. We consider the thin
disc to consist of all stars that lie within the black lines around this
ridge. Less clear is the extent of the thick disc at $\feh\gta-0.6$. Guidance
is provided by plotting in green the locations of those stars in the
realisation that satisfy the kinematic selection criteria of \cite{Bensby03}
to belong to the thick disc. A few of these stars lie in the region reserved
for the thin disc; this phenomenon illustrates the inability of any kinematic
selection criteria to separate cleanly the thin and thick discs -- see
\S\ref{sec:kinsel} below. In light of the distribution of green crosses in
\figref{fig:tcuts}, we define the thick disc to consist of all stars that lie
either above the horizontal line at $\ofe=0.56$ or to the right of the
sloping line, which has the equation
 \begin{equation}
\ofe=0.56-0.55(\feh+0.8) \ .
\end{equation}
 
The lower panel of \figref{fig:tcuts} shows the chemical compositions of
stars in three large observational programs. These studies used different
selection criteria -- \cite{Bensby05} kinematically selected for thick-disc
stars, while \cite{Gilli06} studied stars with planets, so their stars are
all metal-rich.

\begin{figure}
\centerline{\epsfig{file=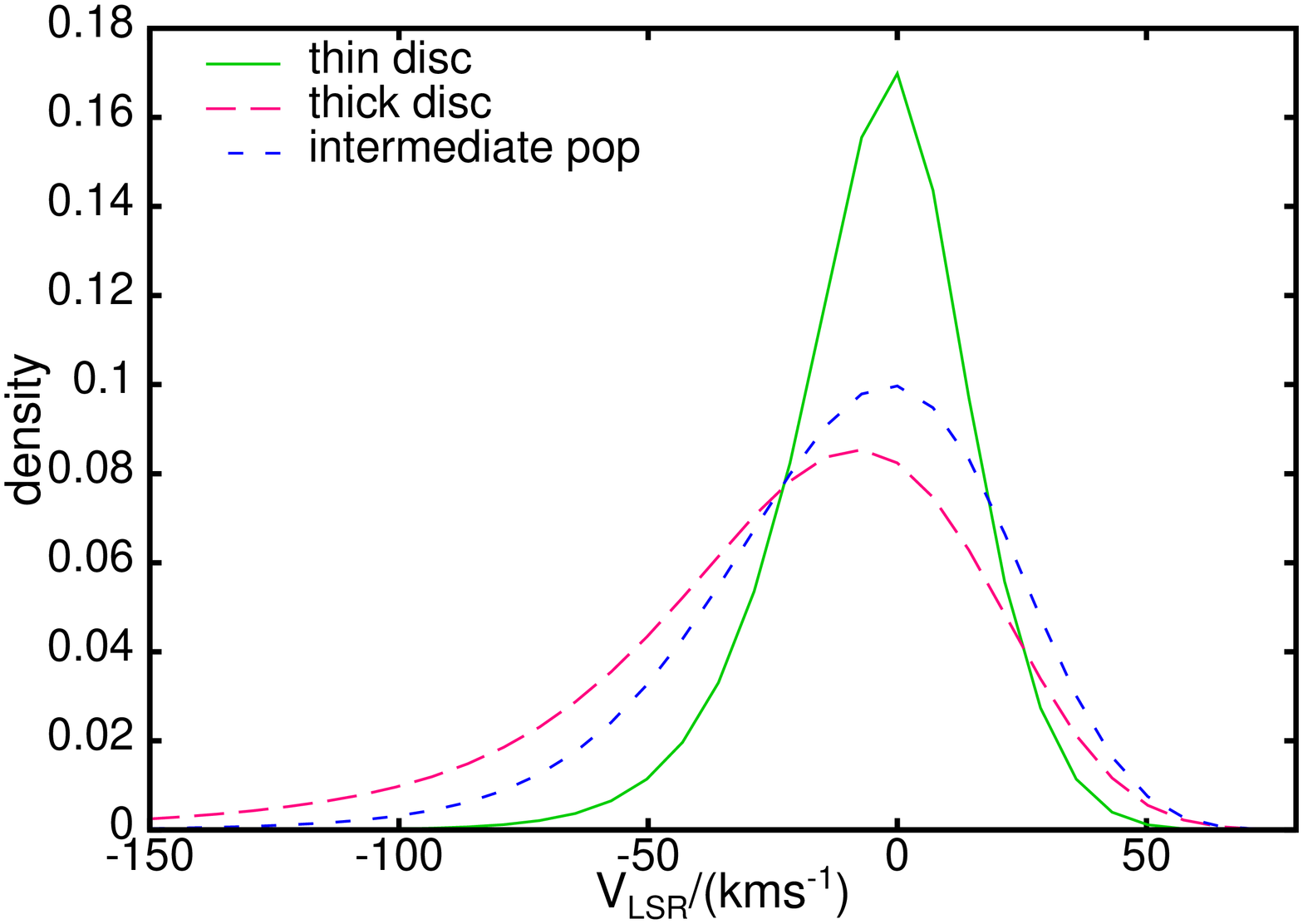,angle=-90,width=.9\hsize}}
\centerline{\epsfig{file=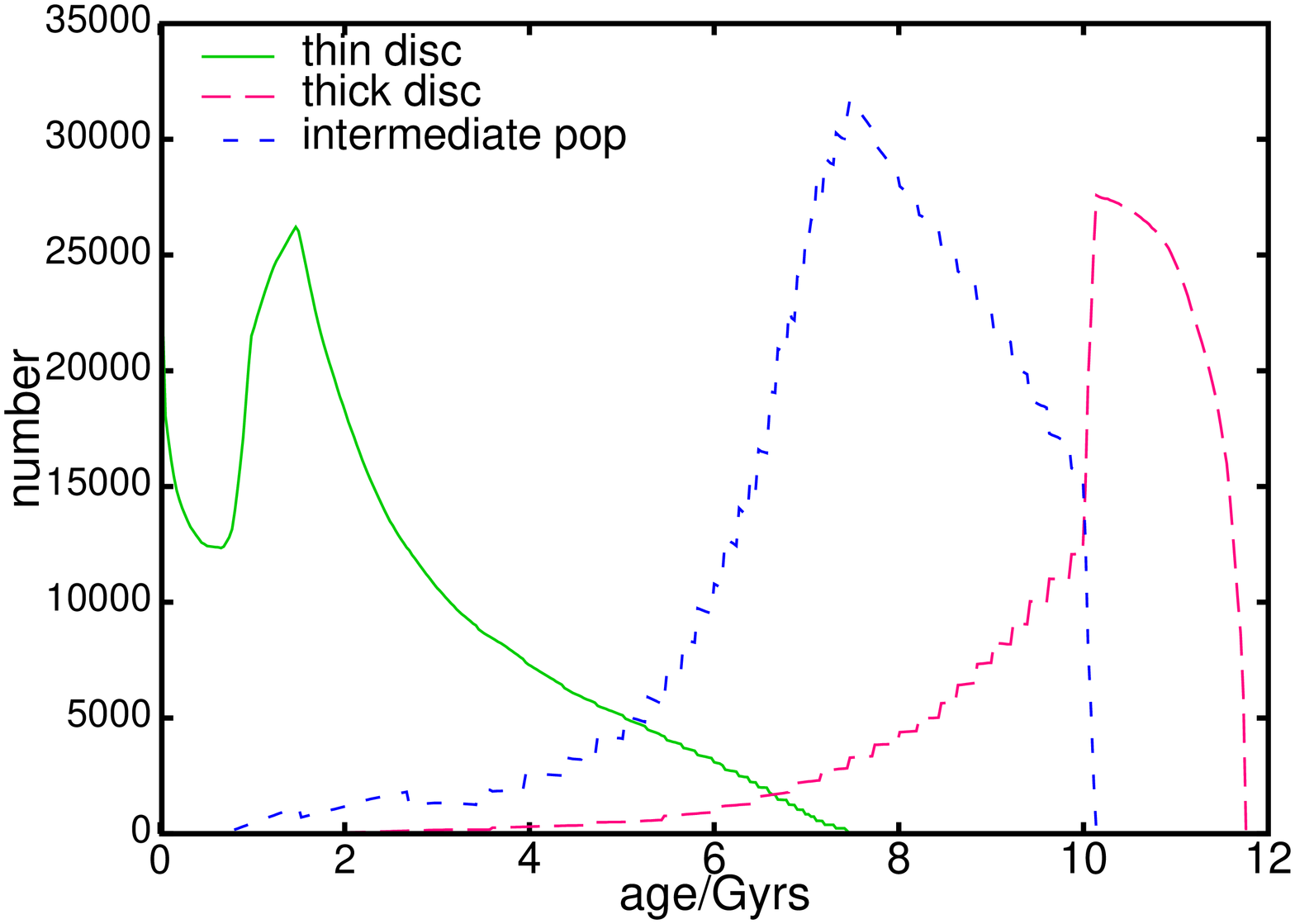,angle=-90,width=.9\hsize}}
\centerline{\epsfig{file=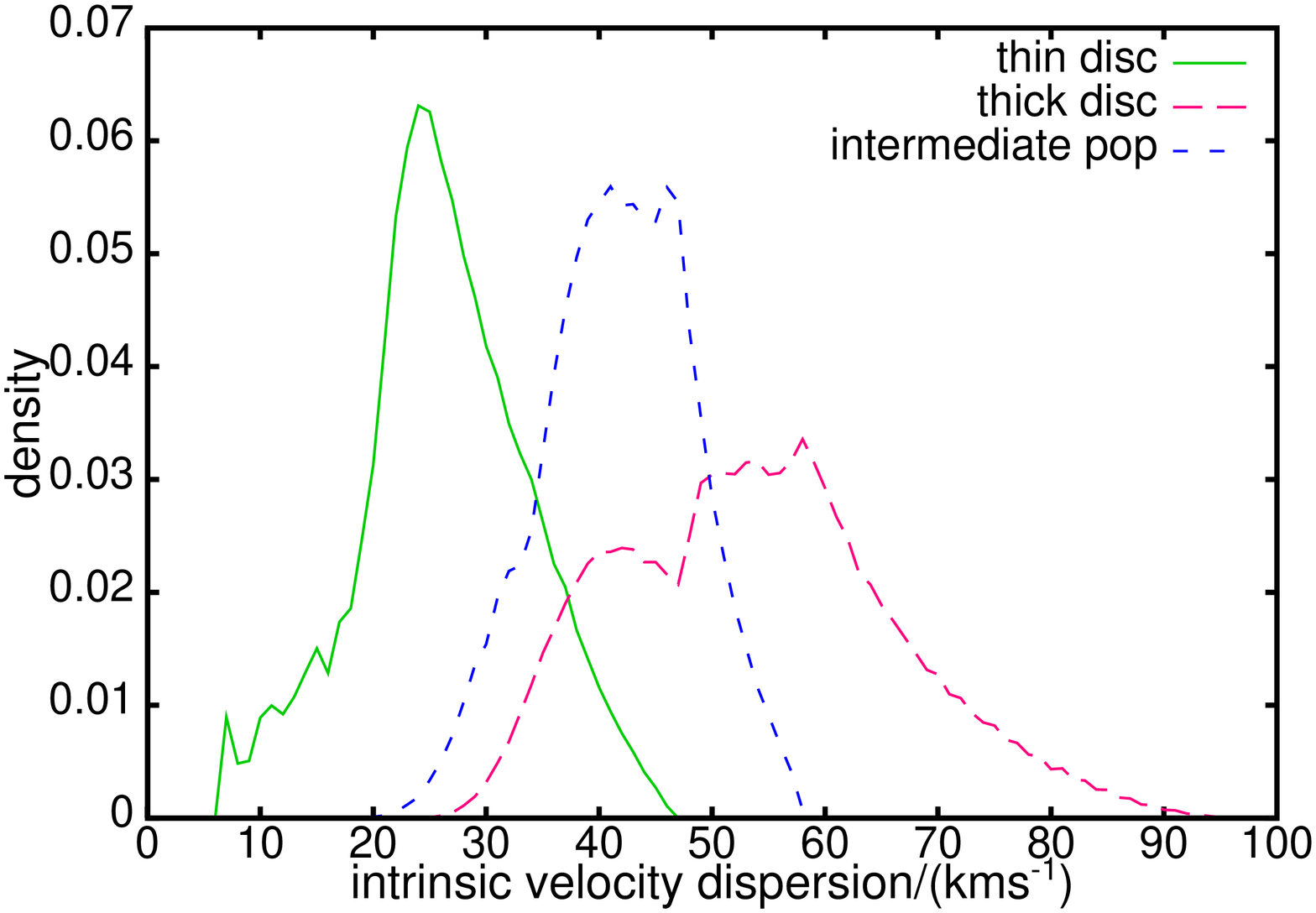,angle=-90,width=.9\hsize}}
 \caption{The top two panels show the distribution of stars from the three
 chemically selected populations in  $V$ velocity and age. The bottom diagram
 shows the distribution of stars by the velocity-dispersion parameter of the
 cohort to which they belong -- see the description in text. The  populations
 are the thin disc (green), the intermediate population (blue) and the thick
 disc (red).  The curve showing the age
distribution of the thin disc has been scaled down by a factor of 10 relative to the
curves for the other two components. In the other panels each curve is separately 
normalised to unity. The steps in the age distribution are artifacts arising
from  the model's radial resolution ($0.25\kpc$); a step is produced as an
individual ring passes over the selection criterion.
\label{fig:VI}}
\end{figure}

The top and centre panels of \figref{fig:VI} show the distributions of
rotation velocity (top) and age (centre) within the thin disc (green), the
thick disc (red) and the intermediate population (blue) when the local
stellar population is divided in this way.  In the top panel the thick disc
stands out for the extent to which its $V$-distribution extends to low $V$.
However, its peak lags circular rotation by only $\sim 10\kms$ because it has
a significant extension to $V>0$. On account of its long tail, the average
asymmetric drift of the thick disc is $\sim 22.5\kms$, which is slower than
that of the thin disc by $\sim 18 \kms$. The intermediate population is much
more symmetrically distributed in $V$ and, like the $V$-distribution of the
thin disc, peaks near $V=0$ with an average drift of $\sim 10\kms$. Note that
these velocities are relative to the local standard of rest (LSR), rather than the
Sun, which is rotating faster than the LSR by $\sim5\kms$ \citep[cf.][]{DehnenB}. Hence relative to
the Sun, the asymmetric drift of the thin disc is $\sim10\kms$.

\cite{Haywood08} showed that the population with moderate $\alpha$
enhancement is a superposition of stars that either combine lower metallicity
(at fixed $\afe$) with fast rotation, or higher metallicity with lower
rotation. The former sub-group bear a clear outer-disc signature, while the
latter sub-group one associates with the thick disc by virtue of their slow
rotation.  Higher metallicities at given $\afe$ point to a flatter trajectory
in the $(\afe, \feh)$ plane of the relevant ISM, i.e. to faster
metal-enrichment before SNIa started to explode. Such fast enrichment is to
be expected in the dense inner disc.  Thus the structure found by
\cite{Haywood08} is an inevitable consequence of chemical evolution in the
presence of radial mixing.

\begin{figure}
\centerline{\epsfig{file=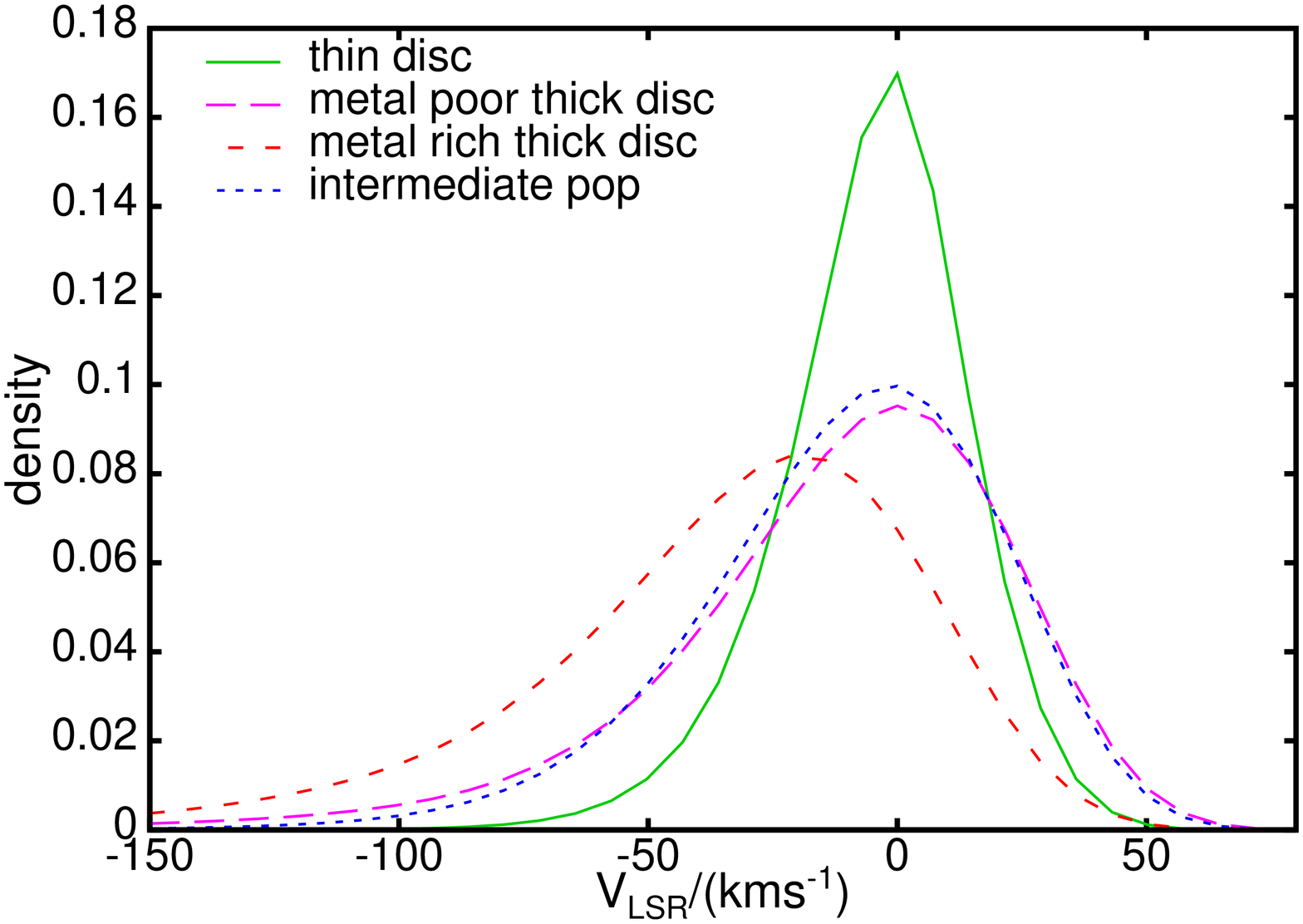,angle=-90,width=.9\hsize}}
\centerline{\epsfig{file=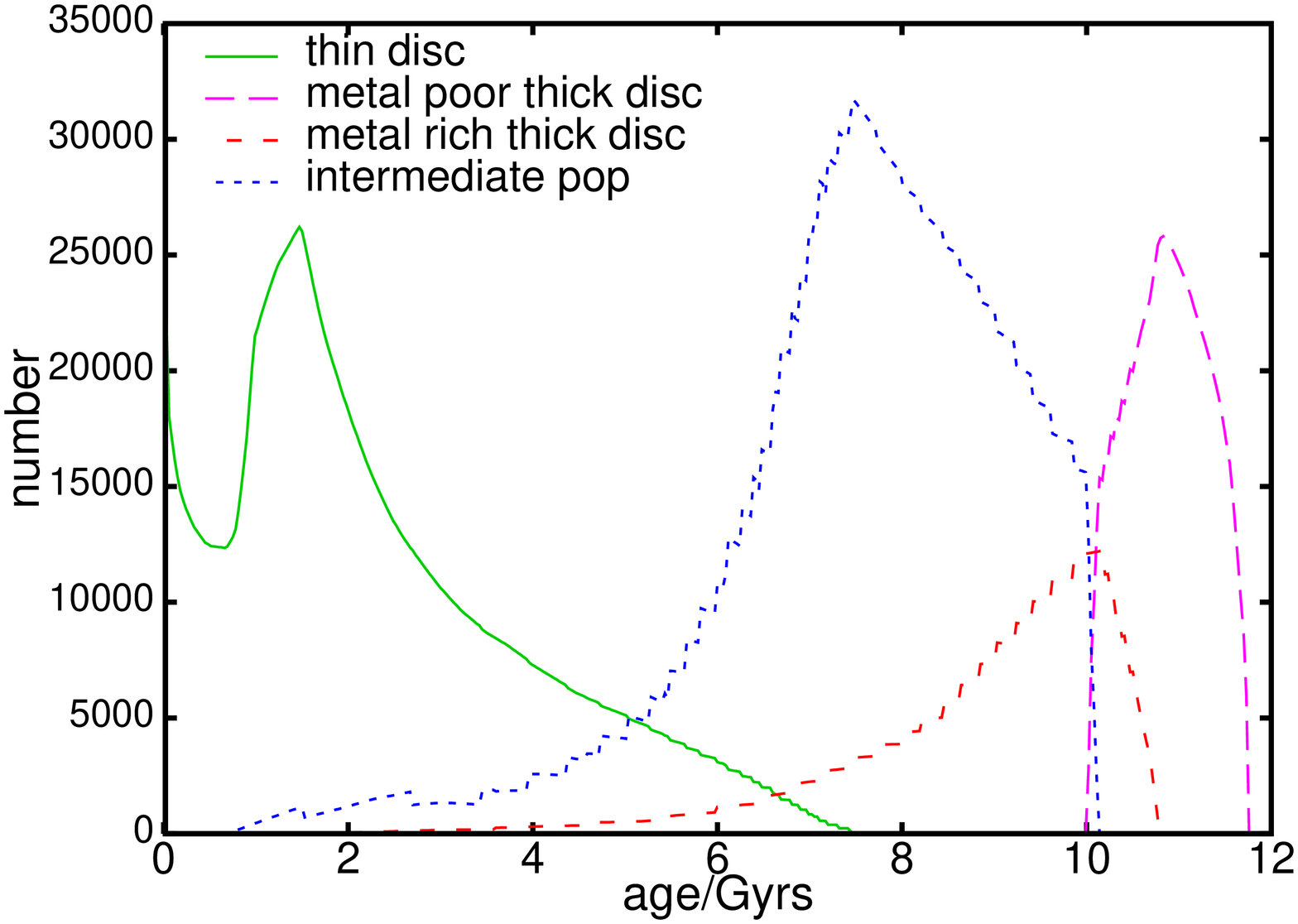,angle=-90,width=.9\hsize}}
\centerline{\epsfig{file=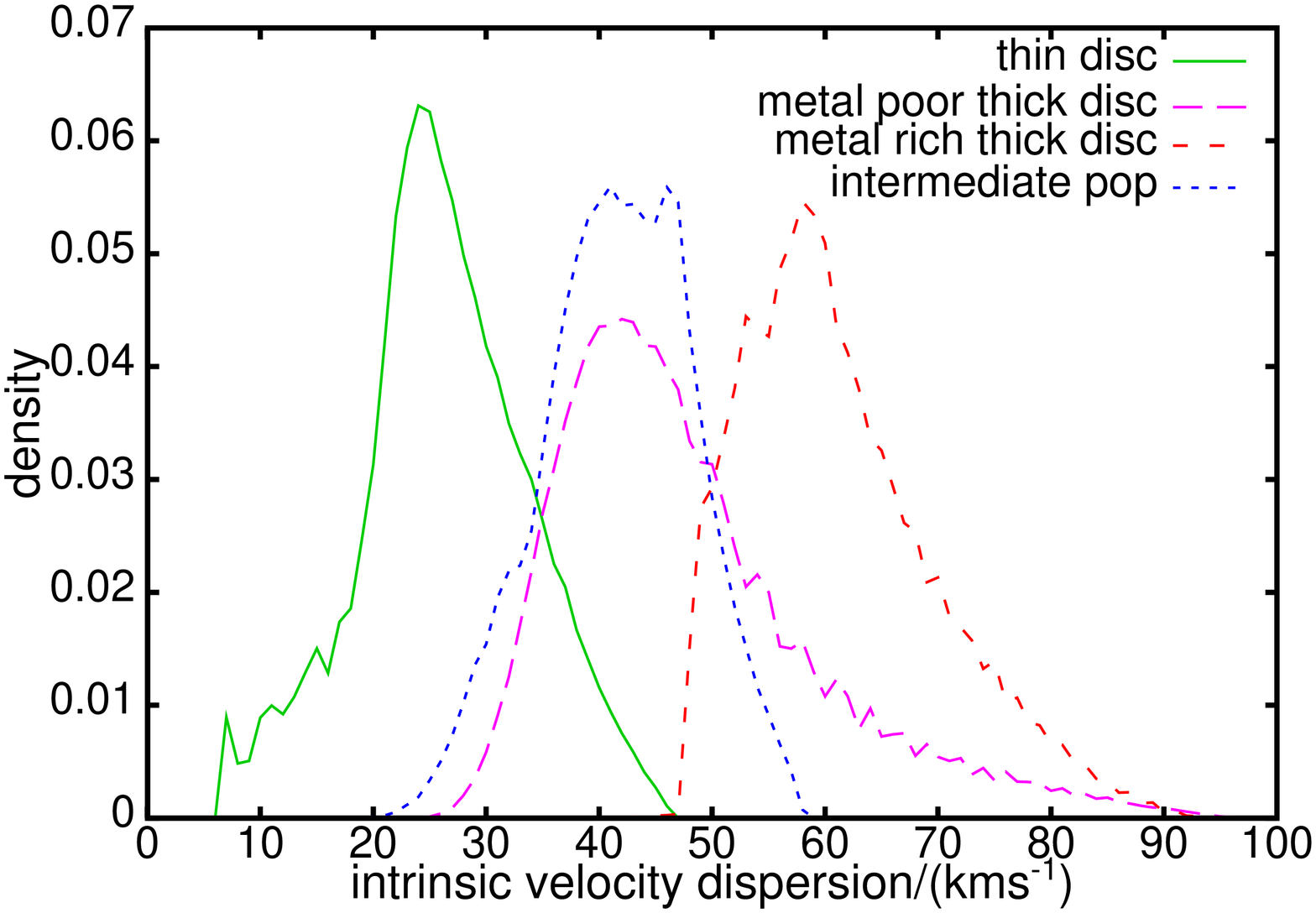,angle=-90,width=.9\hsize}}
 \caption{As \figref{fig:VI} but with the thick disc split into its
metal-weak (purple) and -rich (red) parts, the latter being defined to
comprise
$\feh>-0.8$.\label{fig:VII}}
\end{figure}

The middle panel of \figref{fig:VI} shows that essentially all thick-disc
stars are older than $6\Gyr$. Most stars of the intermediate population are
also this old, but whereas the modal age of the thick disc exceeds $10\Gyr$,
no stars in the intermediate population are older than $10\Gyr$.  The sharp
rise in the number of thick-disc stars at $\sim10\Gyr$, just where the number
of stars in the intermediate population plummets, is very striking. The
purple
curve in the central panel of \figref{fig:VII} clarifies the cause of this
feature by showing the age distribution of stars that have $\ofe>0.56$ (the
horizontal boundary in \figref{fig:tcuts}) and $\feh<-0.8$. We see that all
these metal-poor, highly
$\alpha$-enhanced stars are older than $10\Gyr$, so the significance of a
$10\Gyr$ age is that older stars formed before SNIa started to enrich the ISM
with iron. The purple curve in the top panel of \figref{fig:VII} shows that
the modal rotation velocity of these high-$\alpha$ stars is not far from
circular. That is, the metal-poor thick disc has a smaller asymmetric drift
than the portion of the thick disc that lies to the right of the division line in
\figref{fig:tcuts} (red curves in \figref{fig:VII}), which we henceforth
refer to as the metal-rich thick disc because all its stars have
$\feh>-0.8$. This metal rich portion has an average asymmetric drift of $\sim
35 \kms$, which is $30 \kms$ below the mean rotation velocity of the
thin disc.

The green curve in the central panel of \figref{fig:VI} shows that all thin-disc
stars are younger than $7\Gyr$ and the rate of their formation appears to
rise rapidly towards a peak at $\sim1.5\Gyr$. In reality the SFR in the disc
was monotonically declining throughout this period, so this apparent rise is entirely a
selection effect. Several factors contribute to the detailed shape of the
thin-disc age distribution in \figref{fig:VI}, including the restriction of
the sample to a volume near the Galactic plane (which disfavours old stars)
and the brightening of stars as they begin to turn off the main sequence
(which accounts for the peak at $\sim1.5\Gyr$).

The bottom panels of Figs. \ref{fig:VI} and \ref{fig:VII} show decompositions
of each population into isothermal cohorts. A decomposition is possible
because in the model each cohort of stars (stars formed at a given time and
place) has a steadily increasing velocity dispersion.  The plotted
decompositions show the distribution of the current velocity dispersions for the
cohorts that make up each population, weighted by the fractional
contribution of the cohort to the population.

In \figref{fig:VI} the isothermal decompositions of the thin disc (green) and
intermediate population (blue) are similar except that the distribution for
the intermediate population is shifted to the right by $\sim20\kms$. The
isothermal decomposition of the thick disc is bimodal. The purple curve in the
bottom panel of \figref{fig:VII} shows that the peak of this curve around $40 \kms$ is
associated with the metal-poor thick disc. So within our model the bulk of the
metal-poor thick disc has smaller velocity dispersions than are found in the
metal-rich thick disc. This chimes with the higher characteristic V velocities
of the metal-poor thick disc in indicating that it is cooler and faster
rotating than the metal-rich thick disc. The tail to high dispersions in the
decomposition of the thick disc is contributed by a small number of very old
stars that were formed at small radii, where the velocity dispersion is
currently large.

By fitting  the model's vertical density profile with the sum of two
exponentials in $|z|$,  SB09 concluded that  in the model a fraction
$f_{\rm thick}\sim0.13$ of solar-neighbourhood stars belong to the thick
disc; it followed that of order one third of the entire disc mass is
contributed by the thick disc.  These numbers were in good agreement with the
conclusions that \cite{Juric08} and \cite{Ivezic08} drew from SDSS counts of
stars $\gta1\kpc$ from the plane.  Using the present chemical decomposition
into thin and thick discs, we find $f_{\rm thick}\sim 0.14$. In principle
this number does not have to agree with the value obtained from the density
profile. It does agree well because at $|z|\gta1\kpc$ the disc is dominated
by stars that have thick-disc  chemistry (\figref{fig:metheights}). 

\begin{figure}
\psfig{file=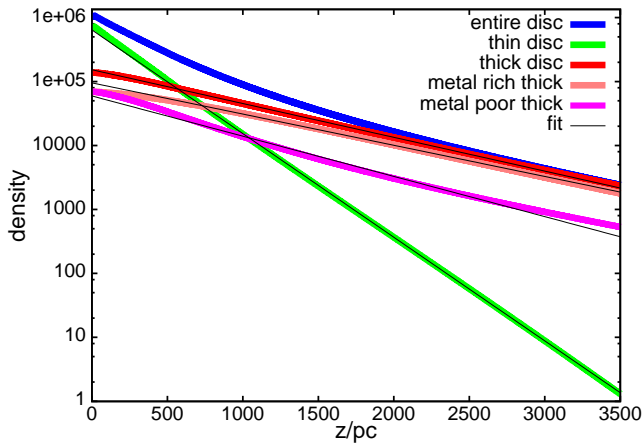,angle=-90,width=\hsize}
\caption{The vertical density profiles of the total disc (blue), thin disc (green), thick
disc (red), the metal poor thick disc component (purple) and the metal
rich thick disc (orange). The thick disc partition at z=0 is $14\%$,
scale heights of the single components are (measured between $z = 400
\pc $ and $z = 3000 \pc$): $h_{thin} = 270 \pc$, $h_{thick}= 820\pc$,
$h_{poor thick} = 690 \pc$ and $h_{rich thick} = 890 \pc$.} \label{fig:vertprof} 
\end{figure}

\figref{fig:vertprof} shows the vertical density profiles of the thin disc
(green), thick disc (red) and the entire disc (blue). Fits of exponentials to
the density profiles yield scale heights of $268\pc$ for the thin disc and
and $822\pc$ for the thick disc. The two components of the latter have scale
heights $690\pc$ for the metal-poor thick disc and $890 \pc$ for the
metal-rich component. All components show more or less exponential profiles.
The metal-poor thick disc has the strongest deviations from an exponential
due to its being a mix of very old stars from all over the disc with
radically
different intrinsic velocity dispersions.  When a sum of exponentials is
fitted to the measured vertical profile of the Galactic disc, good fits can
be obtained with quite a wide range of scale heights on account of a
correlation between the scale heights of the two discs and their
normalisations. The fits above to our individual discs are within the range
of observationally acceptable scale heights \citep[e.g.,][]{Juric08},
consistent with the  thick disc identified by Juric et al.\ being  close to
what we have identified in the model using totally different criteria.

\subsection{Inside-out formation?}

For simplicity the SB09 model does not accelerate the formation of the disc
at small radii relative to large radii, as is required by the popular
``inside-out'' model of disc formation. If the model were adjusted to include
inside-out formation, the main change would be to the metal-poor thin disc,
which would lose parts of its high-$V$ wing. Hence the inside-out scenario could be
put in doubt by demonstrating that the $V$ distribution of the high-$\alpha$ stars
extends significantly to $V>0$, and that many high-$V$ stars have ages in excess of
$10\Gyr$. Further inside-out formation could give rise to some alpha enhanced, relatively
metal-poor stars younger than $\sim 10 \Gyr$ by the later onset of star
formation in outer rings. Neither the thin disc nor the metal-rich thick disc
would be strongly affected by the introduction of inside-out formation.

\section{Kinematic division of the disc}\label{sec:kinsel}

Because it is much easier to measure the velocity of a star than to determine
its chemical composition (particularly its $\alpha$-enhancement), nearly all
analyses select stars kinematically. 
Our model provides an arena in which we can examine the extent to which
kinematically selected samples of each component will be contaminated with
stars from other components.

Samples of local stars such as those of \cite{Venn04} and
\cite{Bensby05} are kinematically divided into thin and thick-disc stars with
the aid of model distribution functions for each component: as described in
\cite{Bensby03}, each star is
assigned to the component whose DF is largest at the star's velocity. Both
DFs are of the type introduced by \cite{Schwarzschild}, namely
 \begin{equation} \label{equ:Bens}
f(U,V,W) = k f_{i}\exp\left(-\frac{{U}^2}{2\sigma_{U}^2}
-\frac{(V-V_{\rm asym})^2}{2\sigma_{V}^2} 
- \frac{W^2}{2\sigma_{W}^2}\right) 
\end{equation}
 where all components are with respect to the Local Standard of Rest,
$k=(2\pi)^{-3/2}(\sigma_U\sigma_V\sigma_W)^{-1}$ is the standard normalisation
constant, $f_{i}$ is the relative weight of the population. The dispersions
$\sigma_i$ assumed for the thick disc are larger than those assumed for the
thin disc, so high-velocity stars tend to be assigned to the thick disc.
Because $V_{\rm asymm}$ is assumed to be $\sim30\kms$ larger for the thick
disc than the thin, stars with lagging rotation velocities and therefore
guiding centres at $R<R_0$ also tend to be assigned to the thick disc. This
effect is reinforced by the fact that the dispersions must increase inwards,
so stars with guiding centers well inside $R_0$ are also likely to be
high-velocity stars.  Consequently the ``thick disc'' stars in \cite{Venn04}
and \cite{Bensby05} tend to belong to the inner disc.  In the context of our
model this fact explains why \cite{Bensby03} found a long tail of ``thick
disc'' stars that have higher $\feh$ at a given $\afe$ than the ``thin-disc''
stars.  

\begin{figure}
\centerline{\epsfig{file=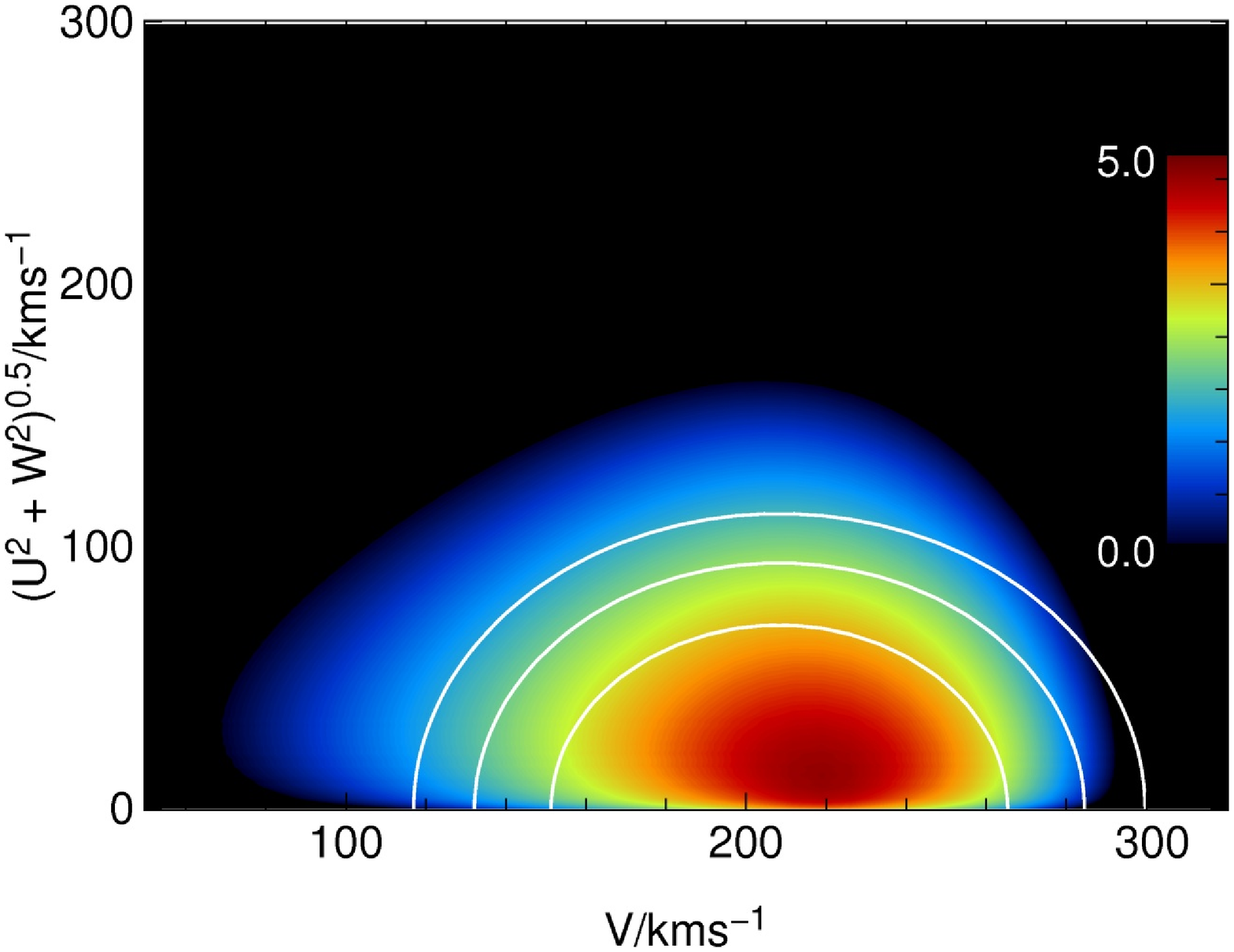,trim=0mm 9mm 0mm
0mm,clip,width=.8\hsize}}
\centerline{\epsfig{file=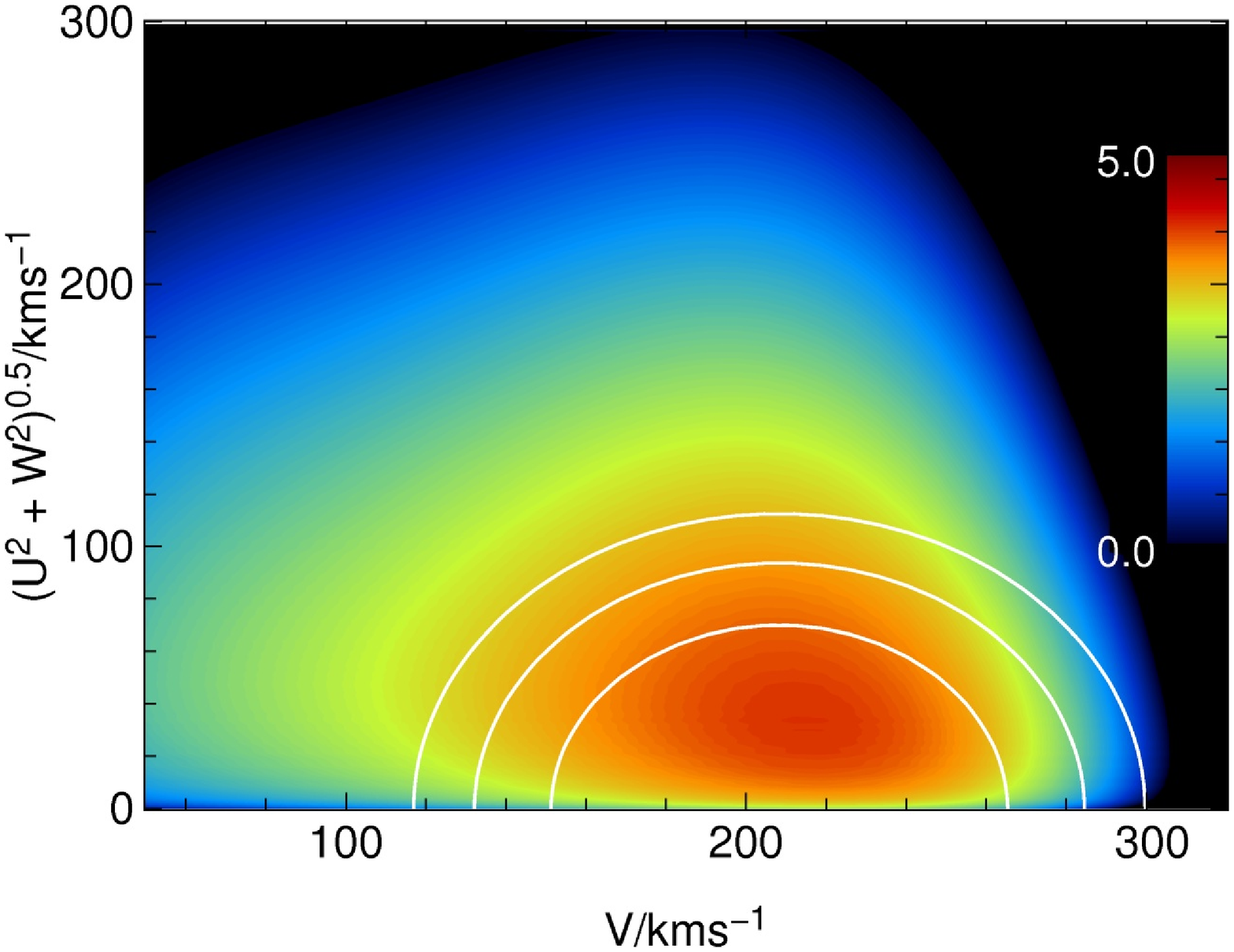,trim=0mm 9mm 0mm
0mm,clip,width=.8\hsize}}
\centerline{\epsfig{file=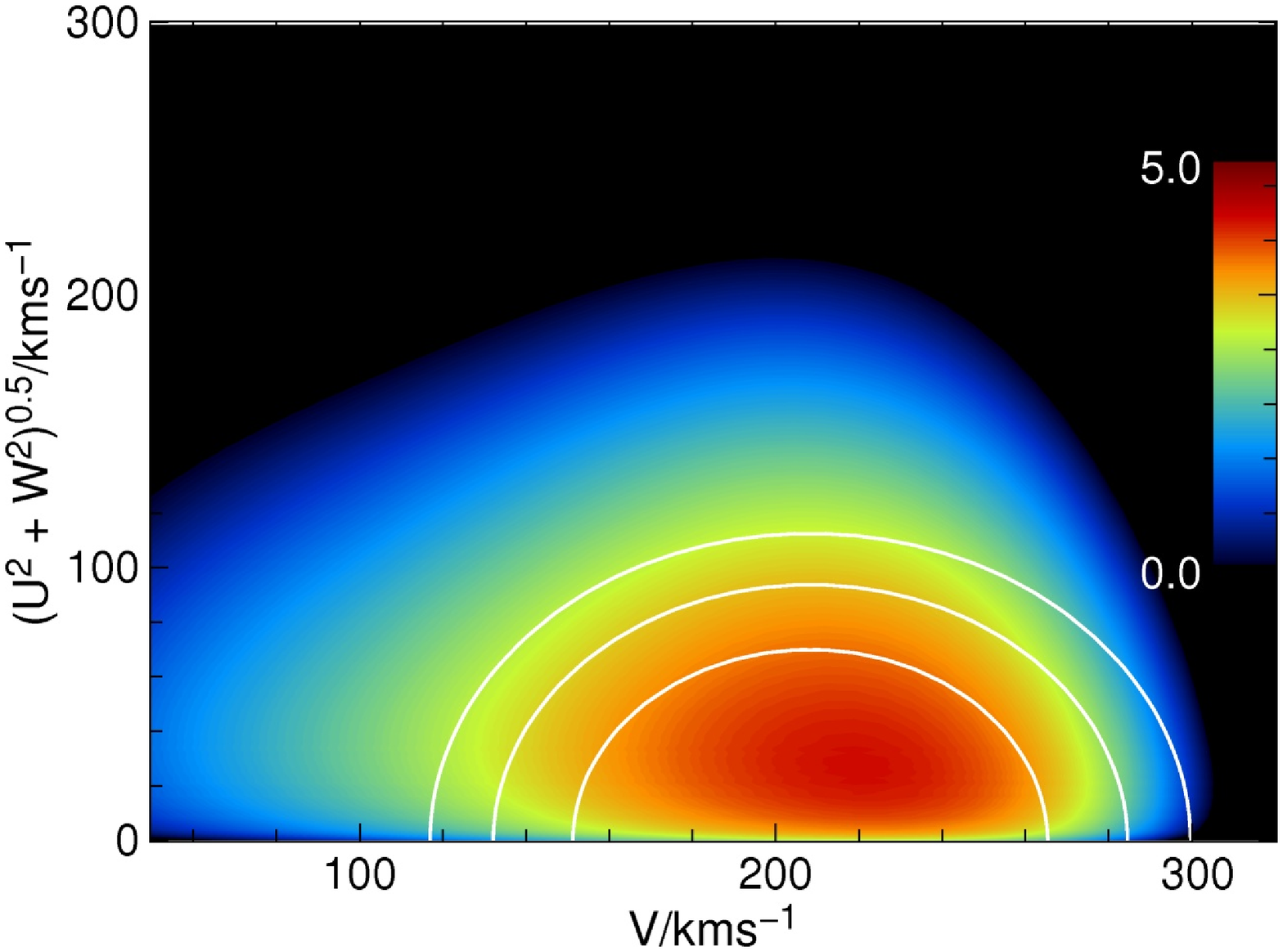,trim=0mm 1mm 0mm
0mm,clip,width=.8\hsize}}
 \caption{From top to bottom Toomre diagrams for the chemically selected
thin-disc, thick-disc and intermediate populations within the model's
solar-neighbourhood. Colour encodes the density of stars and ranges
over 5 orders of magnitude. Shown in white are curves on which the
probability of star belonging to the thick disc by the criteria of Bensby et
al.\ (2003) is constant given $W=0.55U$; on these curves from inwards out the
thick-disc probability is $0.1, 1.0$ and $10.0$ times the thin-probability.
Inside the inner curve stars were deemed to belong to the thin disc, and
outside the outer curve they were assigned to the thick disc.
}\label{fig:toomreI}
\end{figure}

\begin{figure}
\epsfig{file=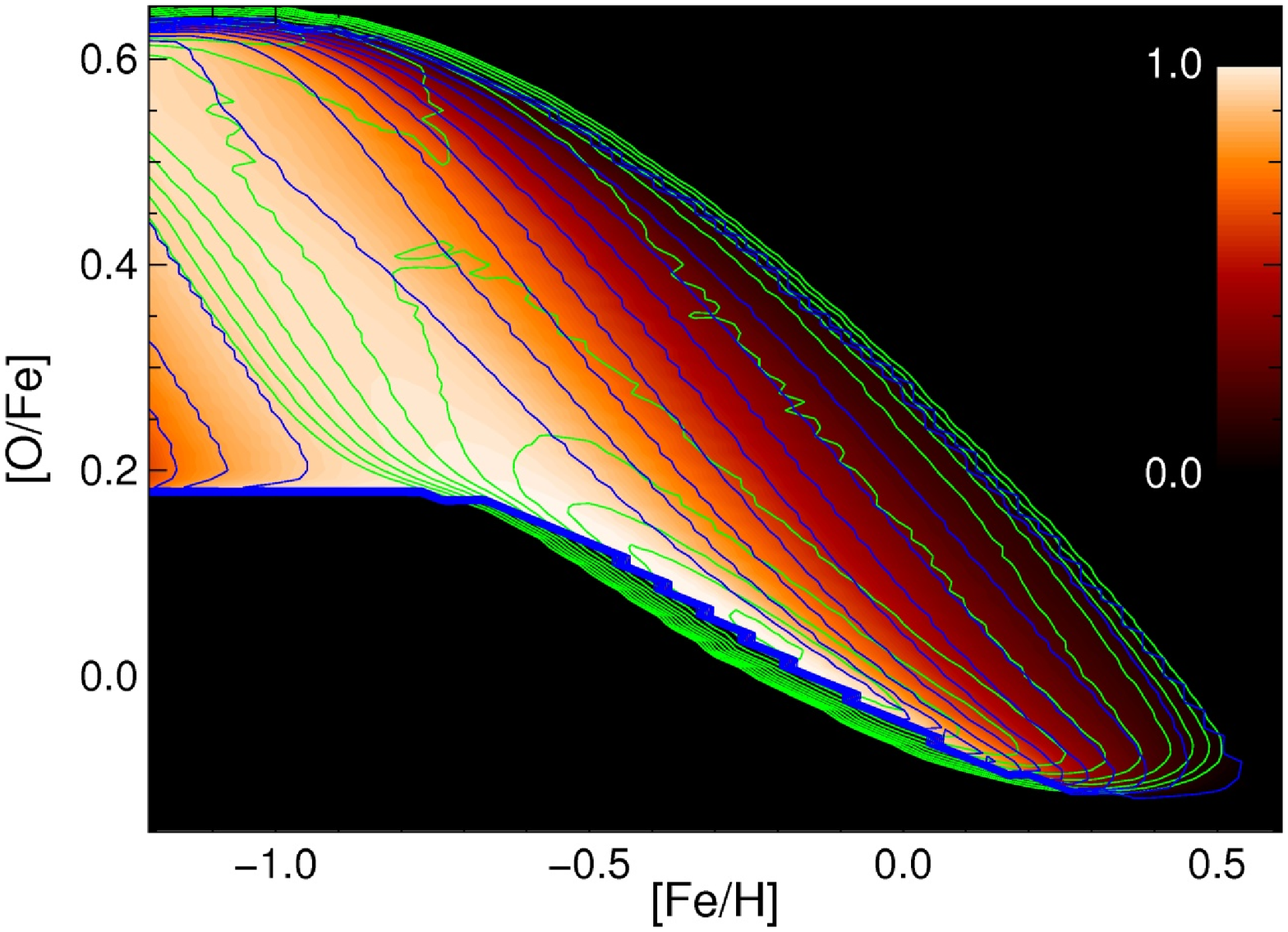,width=\hsize}
\epsfig{file=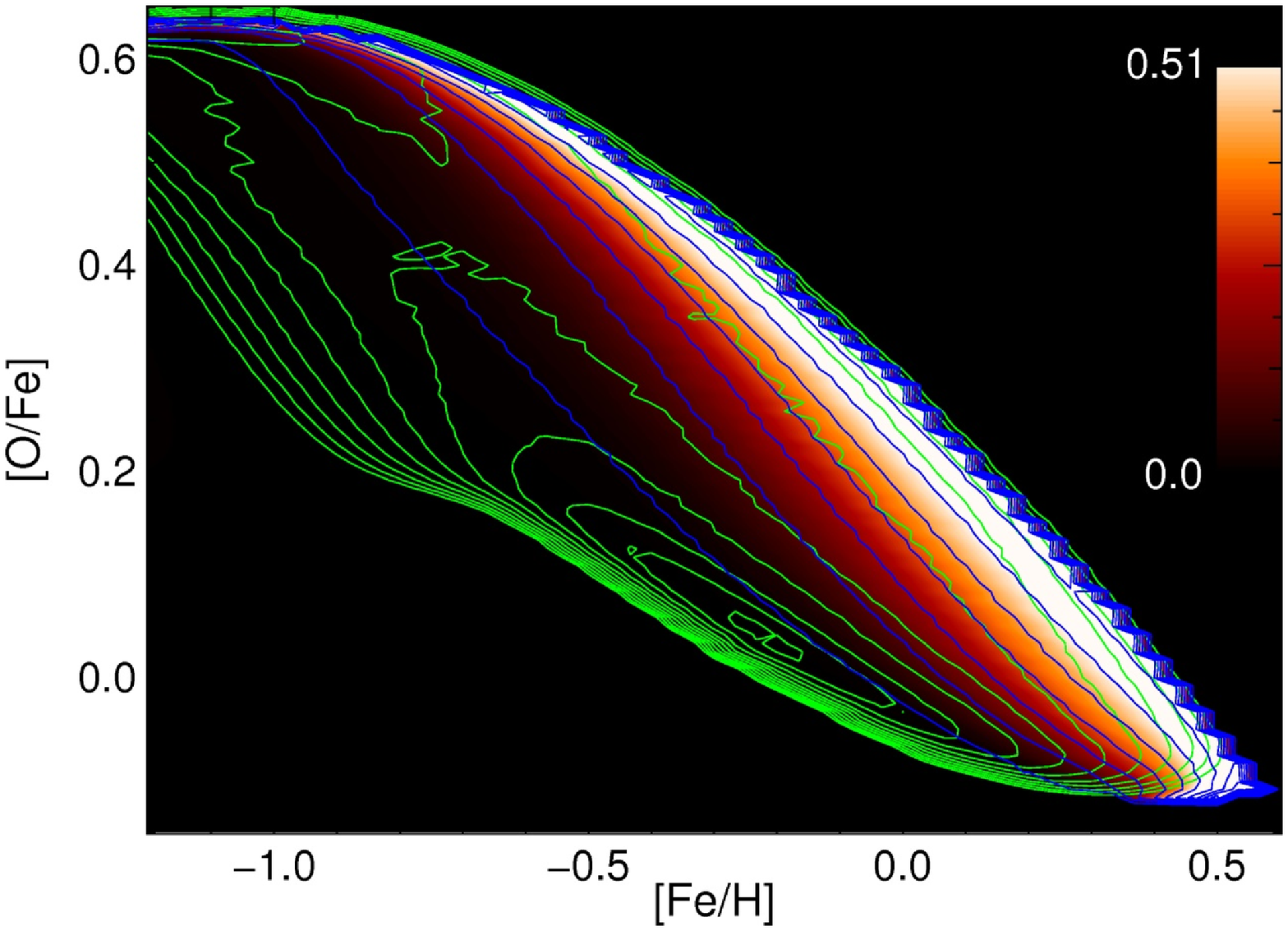,width=\hsize}
 \caption{Selection probabilities using the kinematic selection function of
equation (\ref{equ:Bens}) for the thin (upper panel) and thick disc (lower panel).
Blue contours give lines of same selection probability for a star at a
certain chemical composition with levels running from 0.01 to 0.91 with a 0.1
spacing for the thin disc and from 0.01 to 0.61 with a 0.05 spacing for the
thick disc. Colours encode the selection probability and the green contours
show lines of the model's entire disc population density at a $0.5$ dex
spacing. \label{fig:Bensbysel}}
\end{figure}

\begin{figure}
\epsfig{file=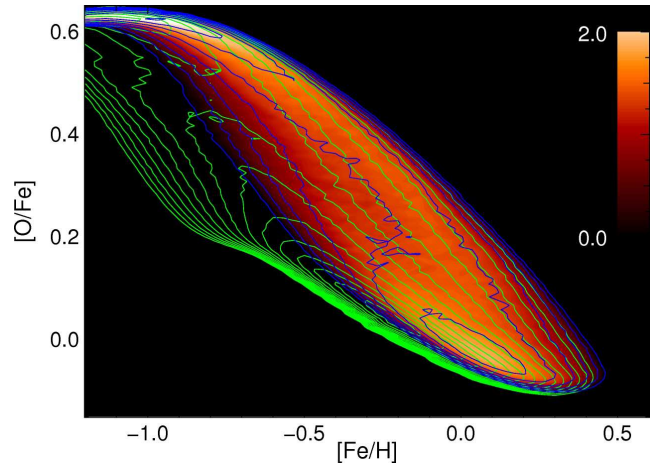,width=\hsize}
 \caption{Green lines show the density contours in the ($\feh, \ofe$) plane of
the entire disc population with a $0.3\dex$ spacing. Colours and blue
contours show the absolute density of stars selected kinematically according
to equation (\ref{equ:Bens}) to the thick disc with $0.3\dex$ contour spacing.
\label{fig:Benf}}
\end{figure}

\begin{figure}
\epsfig{file=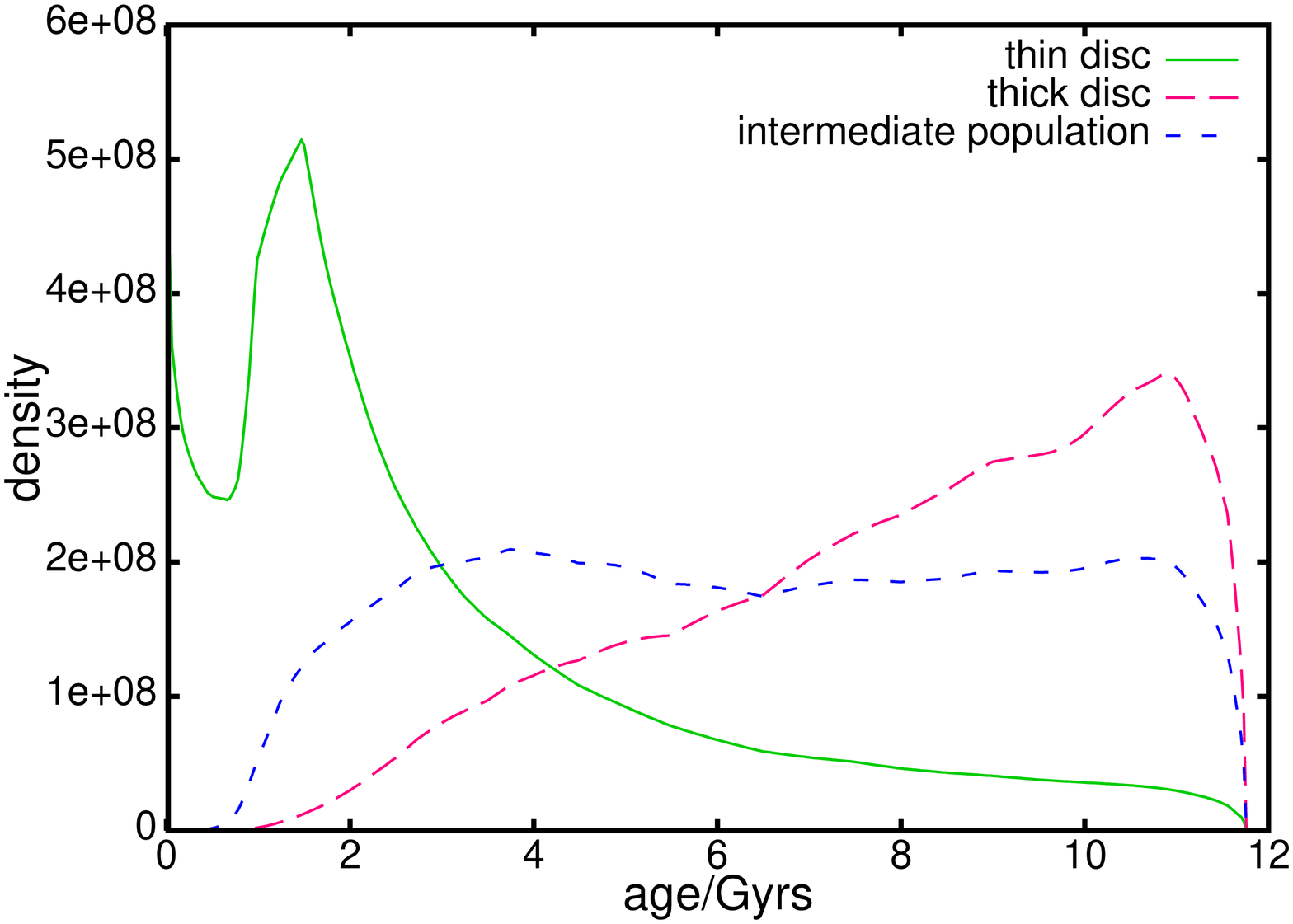,angle=-90,width=\hsize}
\epsfig{file=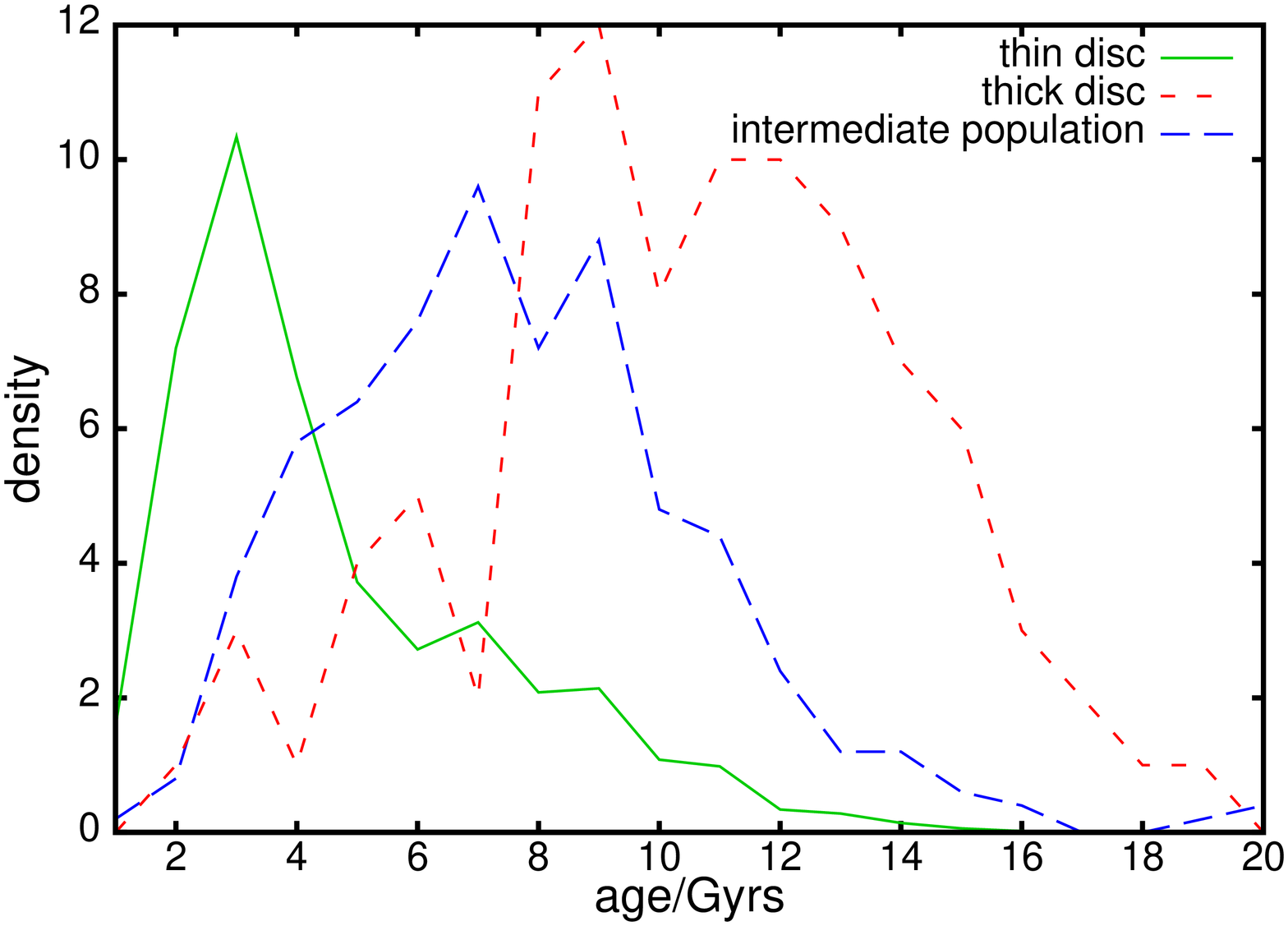,angle=-90,width=\hsize}
 \caption{Upper panel: model age distributions for the three kinematically
selected components in the model.  As in Figures~\ref{fig:VI} and \ref{fig:VII} the curve
for thin disc has been lowered by a factor of 50 and that for the
intermediate population by a factor 5 relative to the curve for the thick
disc. Lower panel: the distribution of measured ages of GCS stars given by
Haywood (2008) with stars kinematically assigned to components using the
Bensby (2003) criteria. Stars with age younger than $0.5 \Gyr$ are not taken
into consideration due to high errors. When the ages published by Holmberg et
al.\ (2007) are used, the lower panel does not change
significantly.\label{fig:Benage}}
\end{figure}

We examine the effectiveness of kinematic selection in two ways. First, in
each panel of \figref{fig:toomreI}, we plot the distribution in a Toomre
diagram of each of the components that we have identified chemically.
Subsequently, in Figs~\ref{fig:Bensbysel}--\ref{fig:Benage} we examine the
distributions in the $(\afe,\feh)$ plane of model stars that have been
kinematically identified as belonging to the thin or thick disc.

\begin{figure}
\epsfig{file=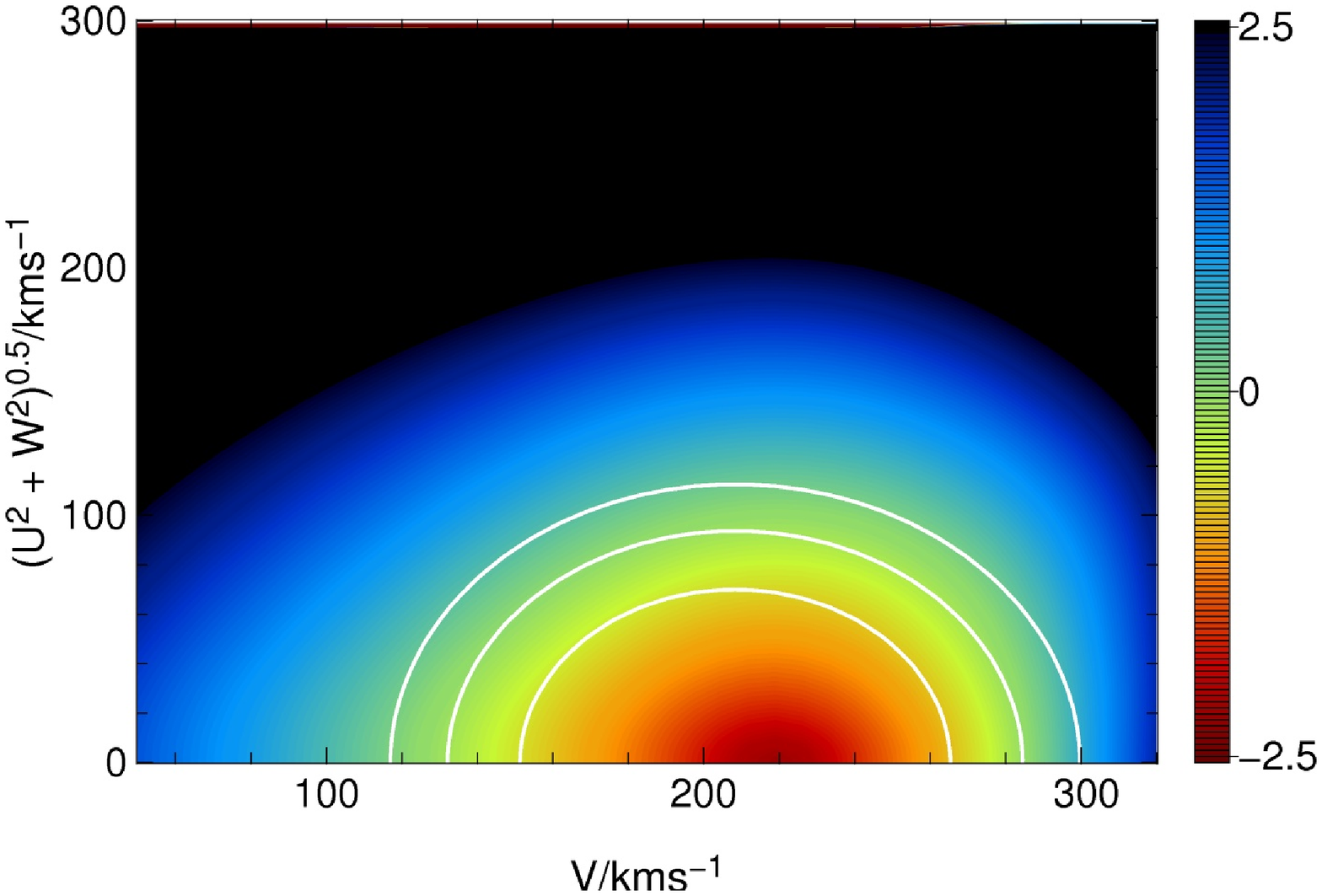,trim=0mm 9mm 0mm 0mm,clip,width=\hsize}
\epsfig{file=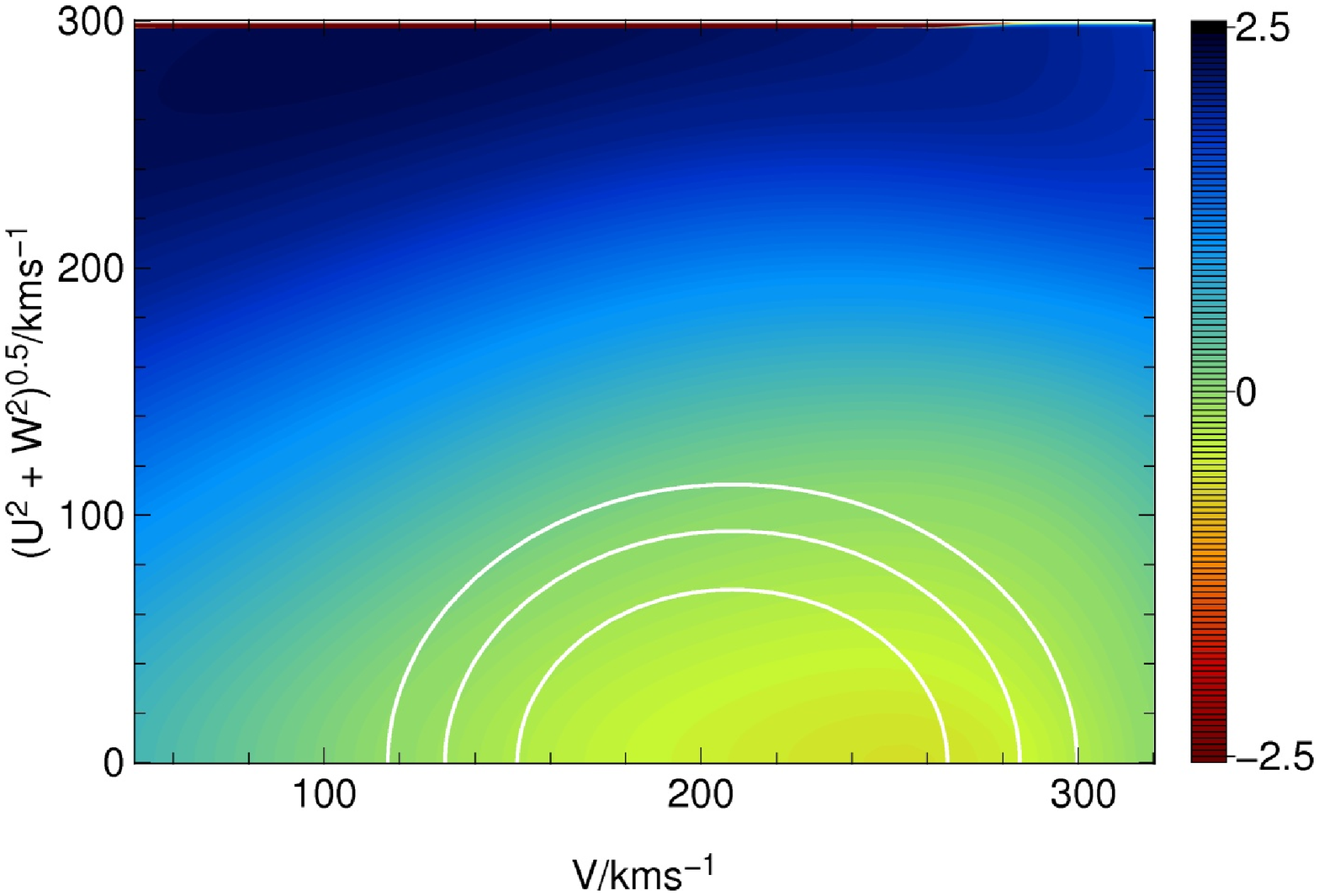,trim=0mm 9mm 0mm 0mm,clip,width=\hsize}
\epsfig{file=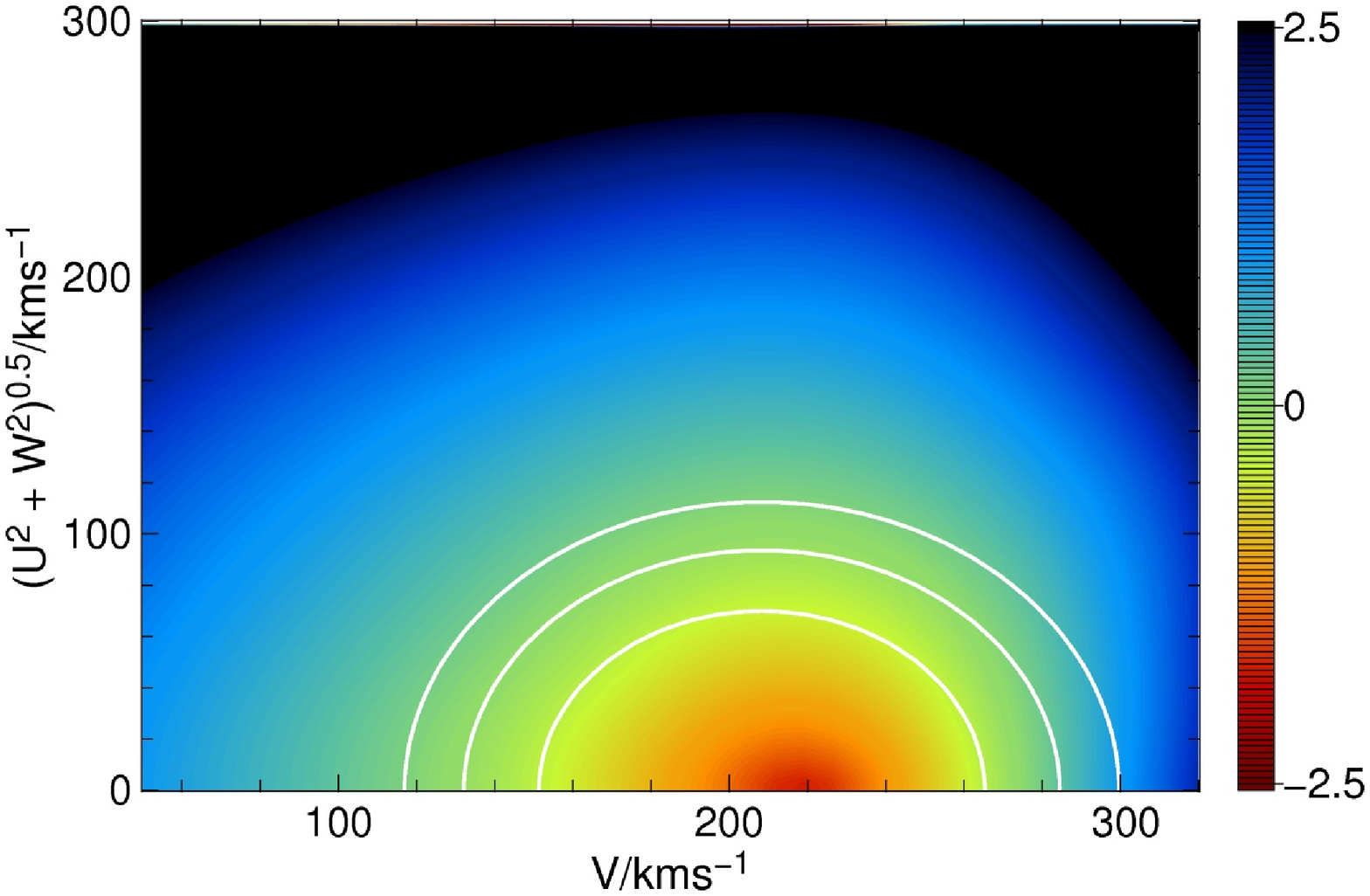,trim=0mm 0mm 0mm 0mm,clip,width=\hsize}
 \caption{From top to bottom the ratio of thick by thin disc, thick disc by
 intermediate disc, and intermediate disc by thin disc stars at each point in
 the Toomre diagram when metal poor and metal rich thick disc are combined.  Colours show the log
of the ratio with values in the range $(-2.5,2.5)$. White contours are the
same as in \figref{fig:toomreI}. \label{fig:toomreIb}}
\end{figure}

\figref{fig:toomreI} shows the Toomre diagrams for the thin, thick and
intermediate components, respectively. In \figref{fig:toomreI} all densities
are separately normalised to unity, while \figref{fig:toomreIb} shows the
density ratios of components in the Toomre diagram. The extensive overlap of
the chemically-selected populations in the Toomre diagram is striking, but a
natural consequence of the approximately Gaussian nature of the distribution
functions of each component, which implies that the density of thick-disc
stars peaks at velocities close to the LSR, which is where the thin disc is
dominant.  Consequently, no kinematic selection of stars from a particular
chemical component can be very clean. This point is underlined by the white
curves in Figs.~\ref{fig:toomreI} and \ref{fig:toomreIb}, which are such that
\cite{Bensby03} classified stars with\footnote{For general values of $W/U$
the \cite{Bensby03} kinematic selection criterion, which is
three-dimensional, cannot be plotted in a Toomre diagram because the latter
is two-dimensional. Hence we choose the approximate ratio of the
dispersion components for our graphs.} $W=0.55U$ as thick-disc if they lay outside the outermost
white curve and thin-disc if they lay inside the innermost curve. The top
panel in \figref{fig:toomreI} shows that this criterion does exclude most
thin-disc stars from a thick-disc sample. However, the upper two panels of
\figref{fig:toomreIb} imply substantial contamination of the thick disc: in
these panels red indicates a region where most stars are not thick disc
stars, yet at lower right red extends significantly beyond the
outermost white curve in both panels. From \figref{fig:toomreI} it is evident
that a slightly cleaner kinematic separation could be obtained if a
non-Gaussian distribution function were used in place of (\ref{equ:Bens}),
but the main problem with kinematic selection is the extensive overlap of
the components in velocity space.

From the centre panel of \figref{fig:toomreI} we see that a large fraction of
the thick disc is also excluded from a kinematically selected sample of
thick-disc stars, and many of the excluded stars will be assigned to the thin
disc because they lie within the region reserved for the thin disc.

\figref{fig:Bensbysel} shows the
probabilities used by \cite{Bensby03} for a star to be assigned to the thin
(upper panel) and thick (lower panel) discs. The probability of being assigned to the thin
disc is large in a broad swath that runs from $\ofe=0.6$ and $\feh=-1.2$ down
to the lower edge of the populated region of the diagram, and then on to solar
$\feh$ and above.  Thus the kinematically selected thin disc includes
high-$\alpha$ stars in contrast to our chemically selected thin disc.  Kinematic
selection does not confine the thin disc to stars near the ridge-line of the
chemical thin disc because the low velocity dispersions and high rotation
velocities characteristic of large radii cause most stars formed at large
radii to be kinematically assigned to the thin disc.  Stars in the more
metal-rich flank of the chemical thin-disc ridge tend to be assigned
partly to the intermediate population, or even (for the highest
metallicities/innermost rings of origin) to the thick disc. 

In \figref{fig:Bensbysel} the probability of a star being assigned to the
thick disc is high along the sloping upper edge of the populated region of
the $(\feh,\afe)$ plane. Consequently, the \cite{Bensby03} kinematic
criterion for being a member of the thick disc does pick stars that belong to
the thick disc by our chemical definition. However, the density of stars
actually assigned to the thick disc by the kinematic criterion, which is
shown in \figref{fig:Benf}, extends below the sloping dashed line in
\figref{fig:tcuts} because the density of stars assigned to the thick-disc is
the product of the assignment probability plotted in \figref{fig:Bensbysel} and
the density of stars in the $(\feh,\afe)$ plane, which declines steeply as the
edge of the populated region is approached.  In fact the ridge  of
kinematically-selected thick-disc stars leaves the upper edge of the
populated region at the
$\alpha$-enhancement turnoff and then runs downwards parallel to
the thin-disc ridge at an offset of $\simeq 0.3\dex$ in $\feh$, just as
reported by \cite{Bensby03} and \cite{Venn04}. Moreover, the zone of
thick-disc stars merges with the thin-disc population at $\feh \simeq 0$,
which was one of the main findings of \cite{Bensby03}.

The upper panel of \figref{fig:Benage} shows the age distributions of the
kinematically-selected components, and should be compared with the middle
panel of Fig.~\ref{fig:VI}. When kinematically selected, the thin disc has a
long tail of very old stars. Conversely, the age distributions of the thick
disc and especially the intermediate population extend to much younger ages
when these components are kinematically selected. It is inevitable that a
kinematically selected thin disc will contain old stars that properly belong
to either the thick disc or the intermediate population because stars with
small velocities relative to the LSR must be assigned to the thin disc, yet
any plausible distribution function for the thick disc will be significantly
non-zero at such velocities. The assignment of young stars to the
intermediate population reflects the red colour in the lower right corner of
the centre panel of \figref{fig:toomreIb} that was discussed above.

The lower panel of \figref{fig:Benage} shows histograms of the ages of GCS
stars in \cite{Haywood08} when stars are assigned to components using the
kinematic criteria of \cite{Bensby03} -- the corresponding histograms of the
ages given by \cite{HolmbergNA} is very similar. Clearly the histograms are
badly distorted by errors in the ages, which scatter stars to unrealistically
large ages, so the horizontal scale of the lower panel is nearly twice that
of the upper panel. None the less, the lower panel seems to be as consistent
with the upper panel as the large errors permit.

The numbers of model stars in the solar neighbourhood that are kinematically assigned
to the three components analysed in the upper panel of \figref{fig:Benage} is
$\hbox{thin}:\hbox{intermediate}:\hbox{thick}=1:0.099:0.0239$. The same ratios for
the observational sample analysed in the lower panel of \figref{fig:Benage}
are $1:0.085:0.029$ in satisfactory agreement with the model's
prediction, but the agreement is actually better than this comparison
suggests, when one accounts for the difference between the selection
functions used to select the observed stars in the lower panel of
\figref{fig:Benage}. If we use the GCS sample without binaries, which our selection function was designed for, the ratios are changed to
$1:0.095:0.029$. When we further remove likely halo stars, the
observational thick disc fraction shrinks to $\sim 0.025$. Indeed, the
fraction of the local column of thick-disc stars that resides 
near the Sun is sensitive to the distribution of $W$ velocities.  The latter
is not tightly constrained because one of the least satisfactory aspects of
the model is the absence of dynamical coupling between horizontal and
vertical motions, which obliges one to make an arbitrary assumption about the
variation with random velocity in the shape of the velocity ellipsoid. It is
worth noting that the model probably has more metal-rich stars high above the
Sun than it should have as a consequence of our assumption that high-velocity
stars are as susceptible to churning as low-velocity stars.

\section{Conclusions}

The thick disc is the Proteus of Galactic physics: depending on which
questions you ask it changes its shape.  Although it has been identified both
by its extended vertical density profile and its distinct kinematics, it is
most usefully characterised chemically, not least because chemical
composition is a permanent feature of a star, whereas distance from the plane
and peculiar velocity are ever-changing properties. Moreover, chemical
composition is intimately connected to the time and place of the star's
birth.

The determination of the chemical composition of large numbers of old
main-sequence stars is feasible only for samples of nearby stars.
Unfortunately, the nearby stars constitute a strongly biased sample of the
whole Galactic disc. It is absolutely essential to interpret the statistics
of the solar neighbourhood in the context of these biases.  We have used our
model Galaxy to explore these biases, and in particular the relationship
between the components one obtains by assigning stars to them on the basis of
their kinematics or their chemistry. A very straightforward conclusion is
that kinematic selection inevitably mis-allocates many stars, both adding old
stars to the thin disc and young stars to the thick disc.

We have shown that our model provides a consistent interpretation of
observations of the solar neighbourhood in which components are identified as
regions of the $(\feh,\afe)$ plane.  Thin-disc stars lie in a narrow ridge of
high density between $\feh\sim-0.65$ and $\feh\sim0.15$ that forms part of
the lower edge of the populated part of the $(\feh,\afe)$ plane. The
metal-rich
thick disc occupies a broader swath of the $(\feh,\afe)$ plane that runs
parallel to the downward-sloping ridge of the thin disc and $\sim0.3$ in
$\ofe$ higher. The metal-rich thick disc extends in $\feh$ from
$\sim-0.9$ to well above 0, where it merges with the thin disc. At its
low-metallicity high-$\alpha$ end, the metal-rich thick disc touches the
metal-poor thick disc, in which $\ofe\simeq0.63\pm0.5$ and $\feh$ goes at
least down to $\sim-1.4$. There is an ``intermediate population'' of stars
that in the  $(\feh,\afe)$ plane lie between the thin and thick discs, but the
density of such stars in the  $(\feh,\afe)$ plane is relatively low. Thus the
two discs are well defined structures.

Thin-disc stars are all younger than $7\Gyr$ and are on fairly circular
orbits. Their values of $\feh$ and rotation velocity $V$ are correlated in
the sense that higher $V$ implies lower $\feh$. The stars of the metal-rich
thick disc are nearly all older than $8\Gyr$. Most are on significantly
non-circular orbits with guiding centres inside $R_0$ and a significant
number have $V<-100\kms$. Specifically, the
metal-rich thick disc can be considered to be a superposition of isothermal
components with radial velocity dispersions between $50$ and $80\kms$,
strongly peaked around $60\kms$. The metal-poor thick disc consists
exclusively of stars older than $10\Gyr$. Its stars have on average more
angular momentum and smaller velocity dispersions than the stars of the
metal-rich thick disc. Among the population of strongly $\alpha$-enhanced
stars there is (cf. \figref{fig:veltra}) an extremely strong negative correlation between $\alpha$ and
$V$.

\cite{Melend08} remarked that the thick disc has similar properties to the
Galactic bulge. This conclusion is natural in the context of our model, in
which the metal-rich thick disc is made up of stars
that have migrated to the Sun from the inner disc, where rapid early star
formation enriched the ISM to significant metallicities before SNIa began to lower
$\afe$.   It is to be expected that many of the stars that formed alongside
the thick-disc stars of the solar neighbourhood are now bulge stars.

Perhaps the most uncertain aspect of the modelling is our assumption that a
star's probability of being ``churned'' to a different angular momentum is
independent of the star's random velocity. Since \cite{SellwoodB} did not
investigate the dependence of churning probability on random velocity, our
assumption could be significantly in error, and it is not implausible that
stars with large random velocities have low churning probabilities. In this
case the thick disc would be less radially mixed than our model predicts. We
will shortly investigate the dependence of churning probability on random
velocity.

Given our model's success in synthesising studies of the Galactic disc into a
coherent picture, it is useful as it stands regardless of the theoretical
considerations that motivated its construction. However, it was not made by
searching an extensive parameter space for a model that would fit the studies
described here. Rather it was made by building a code that combined standard
chemical evolution modelling with a model of dynamical evolution that
reflects the understanding of how spiral structure works that
\cite{SellwoodB} gained from N-body models and analytical
dynamics.  Its two
free parameters were determined from the model's fit to the metallicity
gradient in the ISM and to the metallicity distribution of the GCS stars given
in \cite{Nordstrom04} and \cite{HolmbergNA}. Hence we consider the model's
ability to reproduce the data sets of \cite{Fuhrmann98},
\cite{Bensby03,Bensby05}, \cite{Venn04} \cite{Reddy06}, \cite{Haywood08},
\cite{Juric08} and \cite{Ivezic08} is remarkable and suggests that it has a
sound physical basis. 

The key respect in which the model goes beyond traditional models of chemical
evolution is its inclusion of radial migration by stars and inward flow by
gas. Inward flow of gas is important for the model's success because it
establishes a much steeper metallicity gradient than traditional models
produce. The radial migration of stars is absolutely key, because it
structures the thick disc. Moreover it explains the significant spread in
$\feh$ within the local thin disc, and the correlation that \cite{Haywood08}
identified between $\feh$ and $V$. 

The discovery that the thick disc overlaps the thin disc in $\feh$
presented a challenge to conventional models of chemical evolution because it
implies that there are thick disc stars that have both $\afe$ and $\feh$
higher than in some thin-disc stars. The lower values of $\afe$ in the
thin-disc stars imply earlier times of birth, so how come $\feh$ is lower?
The conventional response to this challenge it to suppose that some violent
event led to a suspension of star formation in the disc, and that during this
hiatus a massive injection of metal-poor gas lowered $\feh$ in the ISM
\citep{Chiappini97,Chiappini01}. An objection to this scenario is that many other
galaxies have thick discs with similar properties to ours
\citep{YoachimDalcanton},
so thick-disc formation should not require special circumstances. We do not
press this argument but would strongly make the point that a model of
chemical evolution that includes only {\it essential\/} physics and has a
single, early, maximum in the  star-formation rate  and a monotonically rising value
of $\feh$ at each radius automatically produces a thick disc with just
the properties observed locally. In fact an $\alpha$-enhanced thick disc
forms because the speed at which $\feh$ rises declines as one moves outwards,
so the value attained by $\feh$ when SNIa start to lower $\afe$ increases inwards.
Spiral structure and the Galactic bar scatter $\alpha$-enhanced  stars formed
at small radii onto more eccentric
and more inclined orbits and even scatters some of them onto orbits of
sufficiently high angular momentum that they are found in the solar
neighbourhood. Readers who want to believe in a violent origin of the thick
disc may continue to do so. But they should be aware that the simplest model
of the chemo-dynamical evolution of the disc that includes all relevant
physics reproduces the data. Hence there is absolutely no {\it evidence\/}
that the thick disc has a violent origin.

\section*{Acknowledgements}
 R.S. acknowledges financial and material support from
Max-Planck-Gesellschaft, from Stiftung Maximilianeum and from Studienstiftung
des Deutschen Volkes.

\label{lastpage}


\begin{thebibliography}{}

\bibitem[Aumer \& Binney(2008)]{AumerB08}
Aumer M., Binney J., 2008, MNRAS, submitted

\bibitem[Bensby et al.(2003)]{Bensby03}
Bensby T., Feltzing S., Lundstr\"om I., 2003, A\&A, 410, 527

\bibitem[Bensby et al.(2005)]{Bensby05}
Bensby T., Feltzing S., Lundstr\"om I., Ilyin I., 2005, A\&A, 433, 185

\bibitem[Chiappini et al.(1997)]{Chiappini97}
Chiappini C., Matteucci F., Gratton R., 1997, ApJ, 477, 765

\bibitem[Chiappini et al.(2001)]{Chiappini01}
Chiappini C., Matteucci F., Romano D., 2001, ApJ, 554, 1044

\bibitem[Colavitti et al(2008)]{Colavitti08}
Colavitti E., Matteucci F., Murante G., 2008, A\&A, 483, 401

\bibitem[Dehnen \& Binney(1998)]{DehnenB}
Dehnen, W. \& Binney, J., 1998, MNRAS, 298, 387

\bibitem[F\"orster et al.(2006)]{Foerster06}
F\"orster F., Wolf C. Podsiadlowski Ph., Han Z., 2006, MNRAS, 368, 1893

\bibitem[Fuhrmann(1998)]{Fuhrmann98}
Fuhrmann K., 1998, A\&A, 338, 161

\bibitem[Gilli et al.(2006)]{Gilli06}
Gilli G., Israelian G., Ecuvillon A., Santos N.C., Mayor M.,  2006, A\&A,
449, 723 

\bibitem[Gilmore \& Reid(1983)]{Gilmore83}
Gilmore G., Reid N., 1983, MNRAS, 202, 1025

\bibitem[Haywood(2008)] {Haywood08}
Haywood M., 2008, MNRAS, 388, 1175

\bibitem[Holmberg et al.(2007)]{HolmbergNA}
Holmberg J., Nordstr\"om B., Andersen J., 2007, A\&A, 475, 519

\bibitem[Ivezic et al.(2008)]{Ivezic08}
Ivezic Z., et al., 2008, ApJ, 684, 287

\bibitem[Jenkins(1992)]{Jenkins92}
Jenkins A., 1992, MNRAS, 257, 620

\bibitem[Juric et al.(2008)]{Juric08}
Juric, M., Ivezi{\'c}, Z., Brooks, A., et al., 2008,  ApJ, 673, 864

\bibitem[Just \& Jahreiss(2007)]{JustJahreiss07}
Just A, Jahreiss H., 2007, arXiv0706.3850

\bibitem[Kennicutt(1998)]{Kennicutt98}
Kennicutt R.C., 1998, ApJ, 498, 541

\bibitem[Lewis \& Freeman(1989)]{LewisFreeman89}
Lewis J.R., Freeman K.C., 1989, AJ, 97, 139L 

\bibitem[Mannucci et al.(2006)]{Mannucci06}
Mannucci F., Della Valle M., Panagia N., 2006, MNRAS, 370, 773

\bibitem[Mel{\'e}ndez et al.(2008)]{Melend08}
Mel{\'e}ndez J., Asplund M., Alves-Brito A., Cunha K., Barbuy B., Bessell M.
S.,  Chiappini C., Freeman K. C., Ramirez I., Smith V. V., Yong D., 2008, A\&A, 484, L21

\bibitem[Nordstr\"om et al.(2004)]{Nordstrom04}
Nordstr\"om B., Mayor M., Andersen J., Holmberg J., Pont F., Jorgensen B.R., 
Olsen E.H., Udry S., Mowlavi N., 2004, A\&A, 418, 989

\bibitem[Reddy et al.(2006)]{Reddy06}
Reddy B.E., Lambert D.L., Allende Prieto C., 2006, MNRAS, 367, 1329

\bibitem[Roskar et al.(2008)]{Roskar1}
Roskar R., Debattista, V.P., Stinson G.S., Quinn T.R., Kaufmann T., Wadsley J., 2008, 675,
L65

\bibitem[Sch\"onrich \& Binney(2009)]{SB08}
Sch\"onrich R., Binney J., 2009, MNRAS, 396, 203

\bibitem[Schwarzschild(1907)]{Schwarzschild}
Schwarzschild K., 1907, G\"ottingen Nachr., 614

\bibitem[Sellwood \& Binney(2002)]{SellwoodB}
Sellwood J.A. \& Binney J., 2001, MNRAS, 336, 785

\bibitem[Venn et al.(2004)]{Venn04} Venn K.A., Irwin M., 
Shetrone M.D., Tout C.A., Hill V., Tolstoy E., 2004, AJ, 128, 1177 

\bibitem[Yoachim \& Dalcanton(2006)]{YoachimDalcanton}
Yoachim P.,  Dalcanton J.J., 2006, AJ, 131, 226

\end{thebibliography}
\end{document}